\documentclass[12pt,prd,preprint,nofootinbib]{revtex4-1}

\usepackage{graphicx}
\usepackage{cancel}
\usepackage{amssymb}
\usepackage{textcomp}
\usepackage{amsmath}
\usepackage{bm}
\usepackage{natbib}
\usepackage{times}
\usepackage{color}
\usepackage{multirow}
\usepackage[pdftex,colorlinks=true, pdfstartview=FitV, linkcolor=blue, citecolor=blue, urlcolor=blue]{hyperref}
\hypersetup{colorlinks=true}

\def\lsim{\mathrel{\raise.3ex\hbox{$<$\kern-.75em\lower1ex\hbox{$\sim$}}}}
\def\gsim{\mathrel{\raise.3ex\hbox{$>$\kern-.75em\lower1ex\hbox{$\sim$}}}}

\def\beq{\begin{equation}}
\def\eeq{\end{equation}}
\def\be{\begin{equation}}
\def\ee{\end{equation}}
\def\bea{\begin{eqnarray}}
\def\eea{\end{eqnarray}}

\def\to{\rightarrow}

\begin{document}

\title{Thermal dark matter implies new physics not far above the weak scale}

\author{Csaba Bal\'azs$^{\bf a}$}

\author{Tong Li$^{\bf a}$}

\author{Jayden L. Newstead$^{\bf b}$}

\affiliation{
$^{\bf a}$  ARC Centre of Excellence for Particle Physics at the Tera-scale, School of Physics, Monash University, Melbourne, Victoria 3800, Australia\\
$^{\bf b}$  Department of Physics, Arizona State University, Tempe, Arizona 85287, USA
}
\date{\today}

\begin{abstract}

In this work we complete a model independent analysis of dark matter constraining its mass and interaction strengths with data from astro- and particle physics experiments.  We use the effective field theory framework to describe interactions of thermal dark matter particles of the following types: real and complex scalars, Dirac and Majorana fermions, and vector bosons.  Using Bayesian inference we calculate posterior probability distributions for the mass and interaction strengths for the various spin particles.  The observationally favoured dark matter particle mass region is 10-100 GeV with effective interactions that have a cut-off at 0.1-1 TeV.  This mostly comes from the requirement that the thermal abundance of dark matter not exceed the observed value.  Thus thermal dark matter coupled with present data implies new physics most likely under 10 TeV.

\end{abstract}

\maketitle

\section{Introduction}


Despite of substantial experimental and theoretical effort the microscopic properties of dark matter particles are unknown.  One problem that makes the extraction of these properties difficult is the plethora of competing theoretical models that provide dark matter candidates.  In principle in each of these models the fundamental properties of dark matter particles can be extracted by contrasting the theoretical predictions with observation.  In practice, however, this task is not feasible due to the sheer number of feasible theoretical models.


To overcome this problem we begin with very general but minimal theoretical assumptions regarding the physics underlying dark matter particles.  We adopt the effective field theory framework which, in principle, contains all specific dark matter models that can be formulated as a quantum field theory \cite{Chen:2013gya}.  For simplicity we augment the Standard Model of particle physics with a single dark matter candidate and assume that all other degrees of freedom are either heavy enough to be integrated out or couple to the observable spectrum with negligible strength \cite{Davidson:2014eia}.  We then calculate dark matter observables and contrast them with experiment.  Using Bayesian inference we can confine the most fundamental properties of dark matter particles, such as their mass and interaction strength with ordinary matter and force particles.


The main price that we pay for relying on effective field theory rather than a specific theoretical model is that our prediction is less informative.  This means that the probability distributions that we extract for the mass and interaction strength of dark matter particles are wider, less peaked, compared to those for a specific model.  In the present work, however, our aim is to show that the single assumption that dark matter is a thermal relic leads to the conclusion that the new physics associated with it is characterized by a mass scale that is not too far from the electroweak scale.  As we will see the effective field theory framework provides us with enough precision for this statement.
Another possible drawback of the effective field theory frameworks is that in some new physics scenarios various assumptions might conspire to change the conclusion of our analysis \cite{Chang:2014tea}.  In the Bayesian spirit, where Occam's razor depletes the probability of increasingly complicated theories, we are willing to take this chance.

\begin{figure}[htb]
\begin{center}
\includegraphics[scale=1,width=0.75\textwidth]{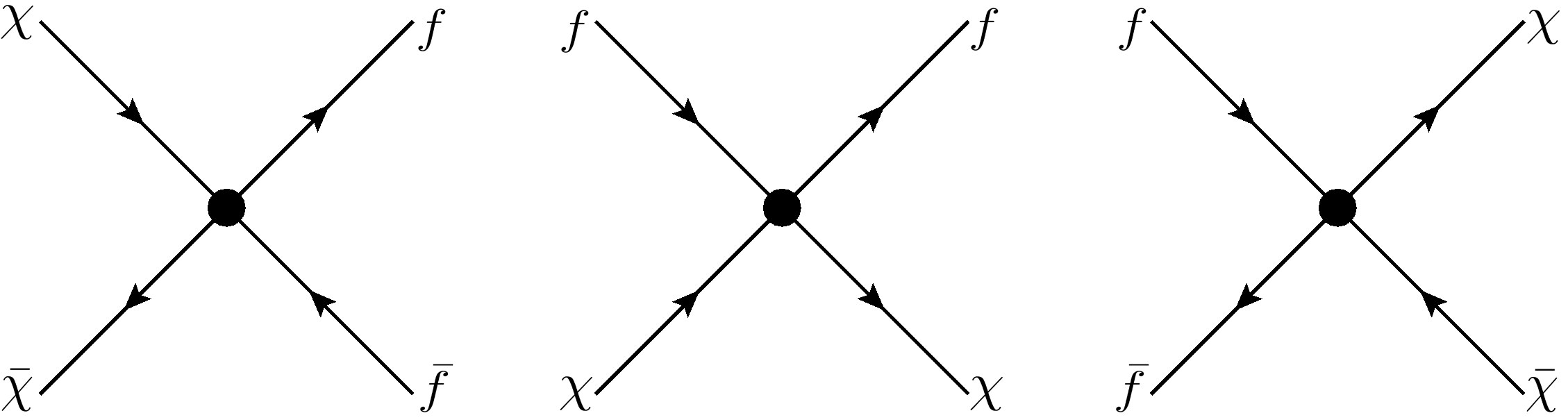}
\end{center}
\caption{Effective interaction between a dark matter particle $\chi$, its anti-particle $\bar\chi$, a standard matter particle $f$, and its anti-matter partner $\bar f$.  The three different orientation of the same diagram shows the three different ways of observing dark matter: indirect detection (left), direct detection (middle) and collider production (right).}
\label{figInteractions}
\end{figure}

Although, to date, there is no non-gravitational evidence for it, dark matter might be observable in three non-gravitational ways.  Based on our knowledge of matter it is expected that dark matter annihilates with its anti-particle or it might decay into standard matter particles.  In this case dark matter annihilation or decay products modify the cosmic ray distribution in our galaxy.  Tantalising deviations from background expectations were found by the Fermi LAT \cite{FermiLAT:2011ab} and AMS \cite{Aguilar:2013qda} collaborations, with conclusive evidence still missing.  Such an observation of dark matter particles would constitute their indirect detection.  In the language of effective field theory the first diagram of Fig.~\ref{figInteractions} shows this possibility for annihilating dark matter.
%
In the early universe the same interaction depleted the (comoving) dark matter density to the level of today \cite{Kumar:2013iva}.

It appears that galaxies rotate faster than estimated based on their ordinary matter content.  The measured rotation rate implies that substantial amount of dark matter is distributed even within our solar system.  Thus, dark matter particles might collide with nuclei within a well shielded detector \cite{Buckley:2013jwa, Panci:2014gga}.  Such collisions may have already been detected by the DAMA \cite{DAMA2010}, CoGeNT \cite{Aalseth:2014jpa}, CRESST \cite{Angloher:2011uu}, and CDMS collaborations \cite{Agnese:2013rvf}.  Detecting dark matter this way is known as direct detection and shown by the second diagram of Fig.~\ref{figInteractions}.  Direct and indirect detection experiments, together with other astrophysical information, provide important constraints on the dark matter mass and its interaction strength with ordinary matter \cite{DelNobile:2013sia, Vecchi:2013iza, Agnese:2014aze, Boehm:2013jpa, Lopez-Honorez:2013cua}.


Based on possible theories of new physics underlying dark matter, such as the WIMP miracle, it is expected that dark matter particles can be produced at high energy and luminosity particle collisions.  The highest energy particle machine, the CERN Large Hadron Collider (LHC), can produce dark matter in proton-proton collisions as shown in the third diagram of Fig.~\ref{figInteractions}.  Since dark matter particles do not leave a trace in the LHC detectors, a signal is searched for in events with dark matter produced in conjunction with a single photon, other weak boson, or a jet.  The analysis of mono-jet plus large missing transverse momentum events at the 8 TeV LHC has pushed the effective interaction scale of Dirac fermion dark matter particles up to 700 GeV and 900 GeV from ATLAS \cite{ATLAS:2012zim} and CMS \cite{CMS:rwa}, respectively.


In this work we confront the experimental limits with the theoretical predictions of effective field theory for the above observables to infer the most probable mass of dark matter particles and their interaction strengths with ordinary matter.  We perform parameter extraction in the context of Bayesian inference.
The rest of the paper is organized as follows.  In Sec.~\ref{EFT}, we recap the effective field theory (EFT) description of dark matter and discuss the dark matter relic density constraint on the mass and interaction scale of various dark matter candidates.  In Sec.~\ref{bayesScan}, we perform a Bayesian analysis of the EFT parameter spaces and show the posterior probability distribution of marginalized dark matter mass, interaction scale and proton-dark matter scattering cross section.  We summarize our results in Sec.~\ref{Sum}.

\section{The Effective Field Theory of Dark Matter}
\label{EFT}

The simplest way to build an effective field theory of dark matter is to introduce a new Standard Model (SM) gauge singlet quantum field, $\chi$.  This field is assumed to be odd under a new parity transformation, the eigenvalues of which are conserved quantum numbers.  Since all the SM fields are assumed to have even parity, $\chi$ is guaranteed to be stable and can only be created or annihilated in pairs.  For completeness we examine five different cases with the $\chi$ field being a real scalar (RS), complex scalar (CS), Dirac fermion (DF), Majorana fermion (MF), and vector boson (VB).  We augment the SM by adding kinetic and mass terms for $\chi$. The interaction Lagrangian containing all Lorentz and gauge invariant operators of dimension-5 for (real or complex) scalar and vector boson, and dimension-6 for Dirac or Majorana fermion particles is scematically given by
\begin{equation}
\mathcal{L}_\chi = \sum_{i,f} C_i \mathcal{O}_{i,f} .
\label{eqnL}
\end{equation}
Here $C_i$ and $\mathcal{O}_{i,f}$ denote a set of coefficients and operators relevant to different structures of $\chi$ interacting with SM fields.  The explicit expressions of $C_i$ and $\mathcal{O}_{i,f}$ are shown in Table~\ref{tableDFoperators}, \ref{tableMFoperators}, \ref{tableVBoperators}, and \ref{tableRCSoperators} for DF, MF, VB, RS and CS dark matter, respectively.  For generality we couple the dark matter field to over all SM fermions $f$ with the exception of the neutrinos.

\subsection{Interaction Operators for Various Spin Cases}

\begin{table}[htb]
\begin{tabular}{|c|c|c|}
\hline
Label & Operator $\mathcal{O}_{i,f}$ & Coefficient $C_i$\\
\hline
D1 & $\bar{\chi}\chi\bar{f} f$ &										 $\frac{m_f}{\Lambda_{D1}^3}$\\[2mm]
D2 & $\bar{\chi}\gamma_5 \chi\bar{f} f$ &								 $\frac{i m_f}{\Lambda_{D2}^3}$\\[2mm]
D3 & $\bar{\chi}\chi\bar{f}\gamma_5 f$ &								 $\frac{i m_f}{\Lambda_{D3}^3}$\\[2mm]
D4 & $\bar{\chi}\gamma_5 \chi\bar{f}\gamma_5 f$ & 					 $\frac{m_f}{\Lambda_{D4}^3}$\\[2mm]
D5 & $\bar{\chi}\gamma^\mu\chi\bar{f}\gamma_\mu f$ & 					 $\frac{1}{\Lambda_{D5}^2}$\\[2mm]
D6 & $\bar{\chi}\gamma^\mu\gamma_5\chi\bar{f}\gamma_\mu f$ & 		 $\frac{i}{\Lambda_{D6}^2}$\\[2mm]
D7 & $\bar{\chi}\gamma^\mu\chi\bar{f}\gamma_\mu\gamma_5 f$ &			 $\frac{i}{\Lambda_{D7}^2}$\\[2mm]
D8 & $\bar{\chi}\gamma^\mu\gamma_5\chi\bar{f}\gamma_\mu\gamma_5 f$ & $\frac{1}{\Lambda_{D8}^2}$\\[2mm]
\hline
\end{tabular}
\caption{The operators and coefficients for a pair of Dirac fermion dark matter coupling to SM fermions, where $\mathcal{O}_{i,f}$ and $C_i$ are used in Eq.~(\ref{eqnL}).}
\label{tableDFoperators}
\end{table}

In Table~\ref{tableDFoperators} with a pair of DF dark matter particles coupling to the SM fermions, operators D1-D4 represent interactions via a heavy scalar mediator, such as the Higgs boson, with varying parity structures.  The inclusion of the fermion mass in the coefficients of these operators prevents flavor violation in the ultraviolet (UV).  Operators D5-D8 represent interactions mediated by a vector particle, again with differing parity alignments.  Further operators could be included to allow for a tensor mediator or to include the possibility of a composite dark matter particle (in this case one could introduce electric/magnetic dipole interactions).  Here we ignore $\chi$ couplings to photon or gluon field strength tensors, as they have one higher dimension and are generated at loop level.

Dimensionless factors could be arbitrarily multiplied to the coefficients of the operators, however, this would drastically increase the size of the parameter space and make our analysis prohibitive.  Since adding the dimensionless factors explicitly would not greatly alter the physics (just the magnitude of the $\Lambda$'s), we instead ignore such factors for all coefficients.  The parameter space is thus more manageable, with a dimensionality of 9: $\Lambda_{D1}-\Lambda_{D8}$ and $m_\chi$.  The above two assumptions are also applied for other spin cases discussed below.

\begin{table}[htb]
\begin{tabular}{|c|c|c|}
\hline
Label & Operator $\mathcal{O}_{i,f}$ & Coefficient $C_i$\\
\hline
M1 & $\bar{\chi}\chi\bar{f} f$ &											$\frac{m_f}{2 \Lambda_{M1}^3}$\\[2mm]
M2 & $\bar{\chi}\gamma_5\chi\bar{f} f$ &									$\frac{i m_f}{2 \Lambda_{M2}^3}$\\[2mm]
M3 & $\bar{\chi}\chi\bar{f}\gamma_5 f$ &									$\frac{i m_f}{2 \Lambda_{M3}^3}$\\[2mm]
M4 & $\bar{\chi}\gamma_5\chi\bar{f}\gamma_5 f$ &							$\frac{i m_f}{2 \Lambda_{M4}^3}$\\[2mm]
M5 & $\bar{\chi}\gamma^\mu\gamma_5\chi\bar{f}\gamma_\mu f$ &				$\frac{1}{2 \Lambda_{M5}^2}$\\[2mm]
M6 & $\bar{\chi}\gamma^\mu\gamma_5\chi\bar{f}\gamma_\mu\gamma_5 f$ &	$\frac{1}{2 \Lambda_{M6}^2}$\\[2mm]
\hline
\end{tabular}
\caption{The operators and coefficients for a pair of Majorana fermion dark matter coupling to SM fermions, where $\mathcal{O}_{i,f}$ and $C_i$ are used in Eq.~(\ref{eqnL}).}
\label{tableMFoperators}
\end{table}

The MF dark matter candidate gains particular attention from the well studied supersymmetric neutralino.
The set of MF dark matter interactions with SM fermions, as shown in Table~\ref{tableMFoperators}, is very similar to the Dirac fermion case.  The difference is due to the Majorana fermion being its own anti-particle, with the consequence that the $\bar{\chi}\gamma^\mu\chi$ bilinear is absent and the general convention includes a factor of $\frac{1}{2}$ in the coefficient.  The same situation happens to RS dark matter candidate compared to CS case as shown in Table~\ref{tableRCSoperators}.

\begin{table}[htb]
\begin{tabular}{|c|c|c|}
\hline
Label & Operator $\mathcal{O}_{i,f}$ & Coefficient $C_i$\\
\hline
V1 & $\chi^{\mu}\chi_{\mu}\bar{f} f$ &							$\frac{m_f}{2 \Lambda_{V1}^2}$\\[2mm]
V2 & $\chi^{\mu}\chi_{\mu}\bar{f} \gamma_5 f$ &					$\frac{i m_f}{2 \Lambda_{V2}^2}$\\[2mm]
V3 & $X^{\mu\nu} X_{\mu\nu} \bar{f} f$ &									$\frac{m_f}{4 \Lambda_{V3}^4}$\\[2mm]
V4 & $X^{\mu\nu} X_{\mu\nu} \bar{f}\gamma_5 f$ &						$\frac{i m_f}{4 \Lambda_{V4}^4}$\\[2mm]
\hline
\end{tabular}
\caption{The operators and coefficients for a pair of vector boson dark matter coupling to SM fermions, where $\mathcal{O}_{i,f}$ and $C_i$ are used in Eq.~(\ref{eqnL}).}
\label{tableVBoperators}
\end{table}

A less thoroughly explored scenario is that of a VB dark matter candidate with couplings to SM fermions shown in Table~\ref{tableVBoperators}. Such a particle may be the gauge boson of a new Abelian gauge symmetry, and in such a case all the SM fields are assumed to be singlets under the same symmetry.

\begin{table}[htb]
\begin{tabular}{|c|c|c|}
\hline
Label & Operator $\mathcal{O}_{i,f}$& Coefficient $C_i$\\
\hline
R1 & $\chi\chi\bar{f} f$ &										 $\frac{m_f}{2 \Lambda_{R1}^2}$\\[2mm]
R2 & $\chi\chi\bar{f}\gamma_5 f$ &							 $\frac{i m_f}{2 \Lambda_{R2}^2}$\\[2mm]
\hline
C1 & $\chi^{\dagger}\chi\bar{f} f$ &										  $\frac{m_f}{\Lambda_{C1}^2}$\\[2mm]
C2 & $\chi^{\dagger}\chi\bar{f}\gamma_5 f$ &								  $\frac{i m_f}{\Lambda_{C2}^2}$\\[2mm]
C3 & $\chi^{\dagger}\partial_{\mu}\chi\bar{f}\gamma^\mu f$ &			  $\frac{1}{\Lambda_{C3}^2}$\\[2mm]
C4 & $\chi^{\dagger}\partial_{\mu}\chi\bar{f}\gamma^\mu\gamma_5 f$ & $\frac{1}{\Lambda_{C4}^2}$\\[2mm]
\hline
\end{tabular}
\caption{The operators and coefficients for a pair of real and complex scalar dark matter coupling to SM fermions, where $\mathcal{O}_{i,f}$ and $C_i$ are used in Eq.~(\ref{eqnL}).}
\label{tableRCSoperators}
\end{table}

The real and complex scalar dark matter scenarios are interesting because these are the simplest extensions to the SM that could solve the DM problem.  The relevant interaction operators are shown in Table~\ref{tableRCSoperators}.

One of the first analyses of effective dark matter interactions was carried out by Beltran et al.~\cite{Beltran:2008}, which was restricted to considering DF dark matter particles. Goodman et al.~explored a comprehensive list of operators which we will draw from~\cite{GoodmanCollider2010,GoodmanCollM2010,GoodmanGamma2010}. While this was a comprehensive analysis, it only considered a single dark matter interaction at a time. Another analysis using gamma ray data was carried out by Cheung et al.~who considered similar operators to Goodman et al.~but instead utilized diffuse gamma ray observations~\cite{Cheung:2011nt}. Cao et al.~considered models of DF, RS and vector bosons, where multiple operators where allowed to contribute~\cite{Cao:2009uw}, though all with the same strength. Other notable analyses include those by Fox et al.~who considered DF dark matter coupled to leptons at LEP~\cite{FoxLEP2011} and to quarks at the Tevatron~\cite{BaiTEV010}. See also the related work by Kopp~\cite{Kopp:2011eu}, produced in collaboration with Fox et al., where constraints from both the Tevatron and LEP are included. Beltran et al.~considered a similar case of dark matter at colliders, but included an outlook for LHC detection prospects~\cite{Beltran:2010}. A more general application of effective theories to identifying new physics at colliders, with emphasis on early results targeted at discovering supersymmetry, was carried out by Alves et al.~\cite{Alves:2011wf}. Taking the UV completion to be at the Planck scale was explored by~\cite{boucenna_2012}, and finally, a thorough compendium of operators at the level of matrix elements can be found in~\cite{kumar_matrix_2013}.

To extend and complement the previous work of others, we generalise the above models and perform a Bayesian analysis on them. We allow for a more general description of reality where more than one operator can contribute, and not necessarily with the same strength. This greatly increases the computational complexity of the problem.  Thus, initially, we restrict the parameter space to a subset of operators that represent common interactions: those mediated by scalars and vectors.

\subsection{Dark Matter Abundance Constraint on the Scale of New Physics}

The Planck satellite measured the dark matter abundance of the universe, in unites of the critical density, to be $\Omega_\chi h^2=0.1196\pm 0.0031$~\cite{Ade:2013zuv}.  For a thermal relic this abundance can be predicted as
\begin{equation}
\Omega_{\chi}h^2 = \frac{8\pi G}{3} s_0 Y_0 m_\chi \simeq \frac{8.33 \times 10^{-12}}{\langle \sigma_{\rm{ann}} v_{\rm{rel}}\rangle_{avg}} .
\label{omegaDF}
\end{equation}
Here $h$ is the present value of Hubble parameter in units of 100 $\rm km/(s\cdot Mpc)$, $G$ is Newton's constant, 
$s_0$ is the present entropy density, and $Y_0$ is the present co-moving number density of the dark matter particles \cite{Beringer:1900zz}.

The thermally averaged annihilation cross section can be calculated in each model via
\begin{equation}
\langle \sigma_{\rm{ann}} v_{\rm{rel}}\rangle_{avg}=\frac{1}{\sqrt{8\pi}}\int_2^{\infty}\sigma_{\chi\chi\to f\bar{f}}(x)x^{3/2}(x^2-4)F_{x_F}(x) dx
\label{sigmaV}
\end{equation}
where the function $F_{x_F}(x)$ and the total annihilation cross section $\sigma_{\chi\chi\to f\bar{f}}$ are given in the Appendix A for various dark matter candidates we consider here.

In general, the interaction cut-off scales in all coefficients cannot be indefinitely large, otherwise, the dark matter annihilation rate would be too slow, which then leads to excessive relic abundance.  There is thus a maximal cut-off scale at which the correct relic density can be satisfied.
To find the approximate upper limit of cut-off scale, we take a universal $\Lambda$ in all operators for various dark matter candidates.  Then we perform the integration in Eq.~(\ref{sigmaV}) with a typical freeze out of $x_F \equiv m_\chi/T = 30$ with $T$ being the freeze out temperature.
Note that, in order to guarantee the validity of the effective field theory framework, the cut-off scale has to be larger than the dark matter mass, i.e. $\Lambda > \frac{m_{\chi}}{2 \pi}$.  We thus set $m_{\chi} = 2 \pi \Lambda$ in the calculation.  This choice also gives the allowed upper limit of $m_\chi$.
The obtained upper limits of universal $\Lambda$ and $m_\chi$ for the various dark matter models are summarized in Table~\ref{tableUpperLimits}.  One can see that the upper limits of $\Lambda$ are at the level of $10^3-10^4$ GeV with the consequent $m_\chi\sim 10^4$ GeV.
The approximate numbers in Table~\ref{tableUpperLimits} agree well with the more precise values we obtain numerically.  During our numerical analysis we determine the preferred $\Lambda$ ranges based on a micrOmegas calculation of the relic abundance ~\cite{micromegas}.

\begin{table}[hbt]
\begin{tabular}{|c|c|c|}
\hline
Model & $m_\chi$ (GeV) & $\Lambda$ (GeV) \\
\hline
DF & $3.0 \times 10^{4}$ & $4.7 \times 10^{3}$ \\
MF & $8.8 \times 10^{4}$ & $1.4 \times 10^{4}$ \\
CS & $4.5 \times 10^{4}$ & $7.1 \times 10^{3}$ \\
RS & $1.1 \times 10^{4}$ & $1.6 \times 10^{3}$ \\
VB & $4.8 \times 10^{4}$ & $7.7 \times 10^{3}$ \\
\hline
\end{tabular}
\caption{The maximal values for the dark matter mass and cut-off scale satisfying relic density constraint in the five considered models.}
\label{tableUpperLimits}
\end{table}

\section{Preferred Mass and Cut-off Regions}
\label{bayesScan}

While mass, spin and interaction strengths are highly sought after fundamental properties of dark matter, Bayesian statistics is the mathematical tool for model parameter extraction.  Based on the available experimental information we can calculate the probability distributions of dark matter mass and cut-off scales using Bayesian inference.
We perform a Bayesian analysis using a multi-modal nested sampling algorithm to scan and extract the parameter space of the dark matter models described in above section.
The masses shown in Table~\ref{tableUpperLimits} are used as upper limit for the parameter scan of $m_\chi$.  We vary cut-off scales for different operators and allow $\Lambda_i$ to be less constrained to account for the possibility of an operator not being present in the full theory.  A value of $3 \times 10^6$ GeV for maximal $\Lambda_i$ was found to sufficiently suppress the presence of extra operators.  The lower limits of $m_\chi$ and $\Lambda$'s were set at 2 GeV.

In our calculation the total annihilation cross sections are computed using CalcHEP~\cite{calchep04}, with the model files generated from LanHEP~\cite{lanhep10}.  Subsequent calculations, including relic density, direct and indirect detection cross sections are implemented with micrOmegas~\cite{micromegas}.  Nested sampling and posterior distribution calculations are performed by Multinest~\cite{multinest08}. Likelihood functions for the relevant experimental constraints are discussed in Appendix B.  The experimental data is drawn from Planck~\cite{Ade:2013zuv} for relic abundance and LUX~\cite{LUX}, CDMSlite~\cite{CDMSlite} and XENON100~\cite{Aprile:2013doa} for direct detection.  Gaussian kernel smoothing is applied before plotting the resultant credible regions of the marginalized posteriors. Note that in some cases the smoothing pushes into disallowed regions, in these cases the credible regions should be thought of as overly conservative. We defer further details of our Bayesian analysis to Appendix B.

\subsection{Posterior Probabilities for $m_\chi$ and $\Lambda_i$}
In this section, we show the posterior probability results for various dark matter candidates based on DM relic density and direct detection constrains.

\subsubsection{Dirac Fermion}

\begin{figure}[htb]
\centering \hspace{9.20mm}
\includegraphics[width=8.07cm,trim=0mm 6.2mm -33mm 0mm,clip=true]{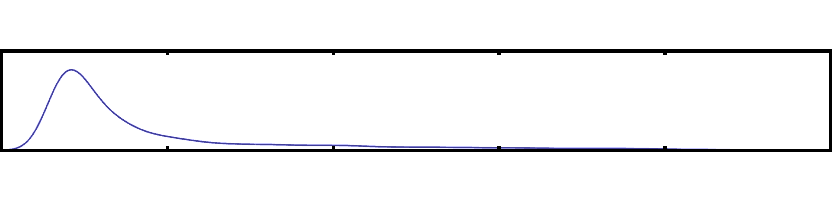} \\
\vspace{-3.9mm}
\includegraphics[width=8cm]{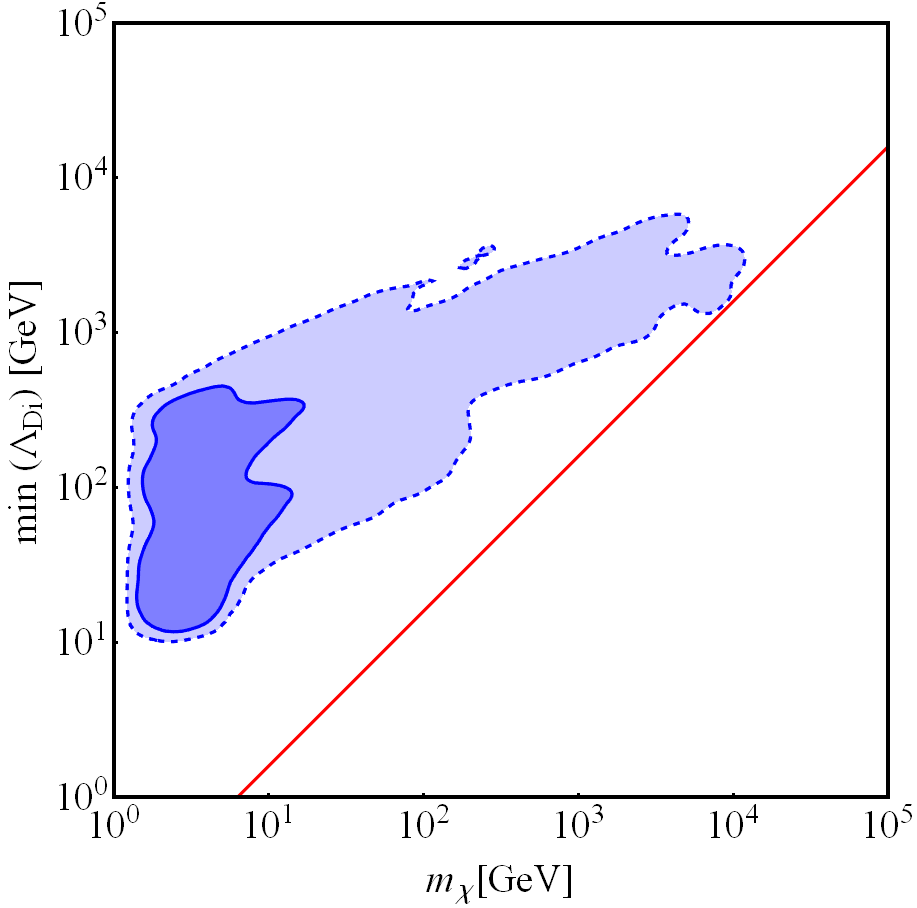} \hspace{-4.6mm}
\includegraphics[width=7.11cm,trim=0mm 10mm -8.mm 0mm,clip=true,angle=-90,origin=r]{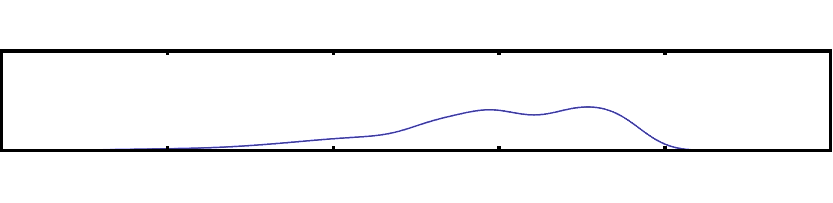}
\caption{Posterior probability distribution marginalized to the minimal $\Lambda$ scale and the mass of the Dirac fermion dark matter particle.  The light and dark regions correspond to 1 and 2$\sigma$ credible regions, respectively. The red line corresponds to $\Lambda = \frac{m_\chi}{2\pi}$.}
\label{figLambdamDF}
\end{figure}

First we show the resulting posterior probability distribution for the Dirac fermion, marginalized to the minimal cut-off scale vs. the dark matter mass in Fig.~\ref{figLambdamDF}.  The minimal cut-off is taken as the smallest value for $\Lambda_i$ in the set of operators.  This is done to capture the dominant operator, contributing the most to the relic abundance or the direct detection cross section, at each point in the multiple dimension space.  We focus on the minimal scale because in this work we are primarily interested in the scale of new physics that is the closest to the electroweak scale.  The results for individual cut-offs are displayed in Appendix C.  The DF dark matter mass is favoured to be less than $100$ GeV in the 1$\sigma$ credible region, with a significant portion below 10 GeV.  The minimal cut-off scale is spread over a wider range, i.e. 10 GeV$<\Lambda<10^3$ GeV.

\begin{figure}[htb]
\centering
\includegraphics[scale=1,width=8cm]{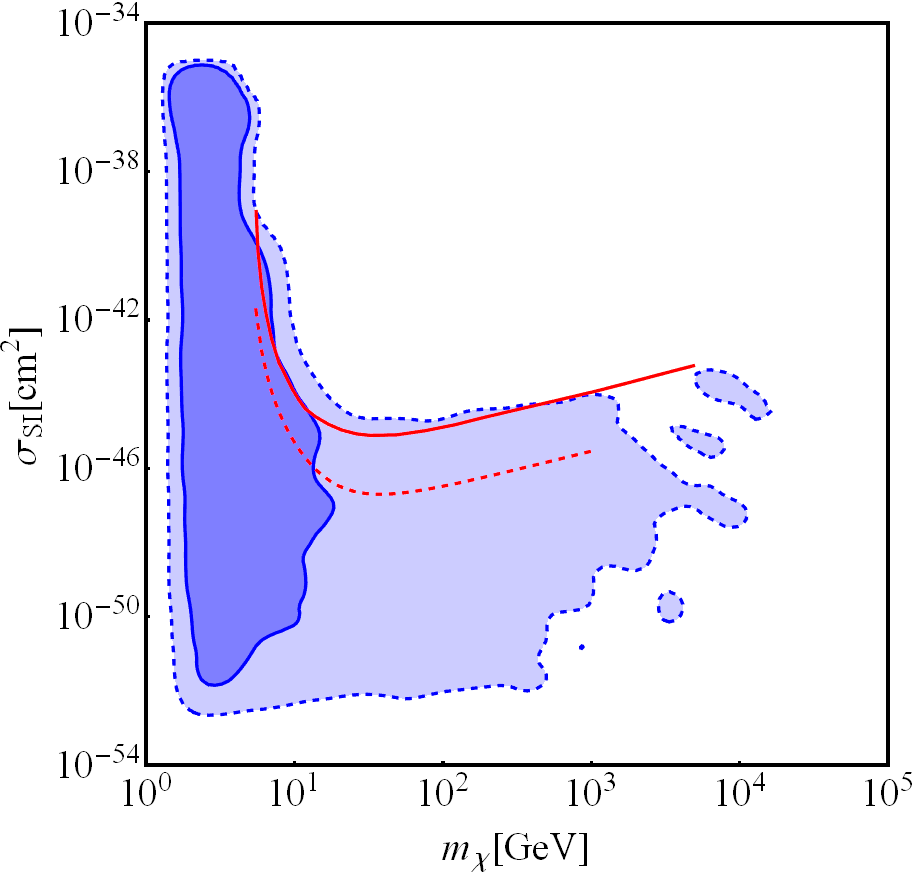}
\includegraphics[scale=1,width=8cm]{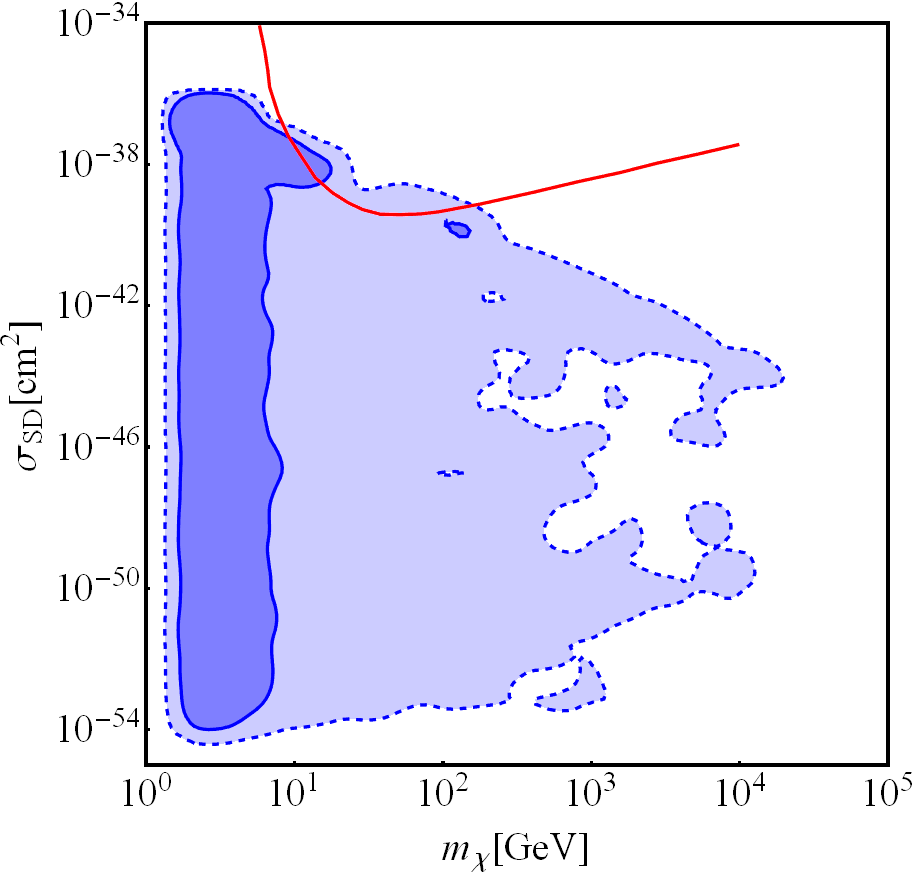}
\caption{Dark matter-proton elastic scattering cross section versus the mass of the Dirac fermion dark matter particle.  The spin-independent (dependent) cross section is shown in the left (right) frame.  The light and dark regions correspond to 1 and 2$\sigma$ credible regions, respectively. LUX (SI) and Xenon100 (SD) 90\% exclusion curves are shown in solid red, projected Xenon1t (SI, 2.6 tonne-years, retrieved via DMtools) limits are in dashed red.}
\label{figSigmamDF}
\end{figure}

The posterior distributions marginalized to the dark matter-nucleon cross sections (spin-independent $\sigma_{SI}$ and spin-dependent $\sigma_{SD}$) as a function of $m_\chi$, are shown in Fig.~\ref{figSigmamDF}.  For DF dark matter, in the non-relativistic and zero momentum transfer limit, the contributions to $\sigma_{SI}$ come from scalar and vector couplings (D1 and D5) and the contribution to $\sigma_{SD}$ is from axial-vector coupling (D8).  The large number of operators in this model allows the direct detection cross sections to take on a wide range of values, such that even the $1\sigma$ credible region spans 14 orders of magnitude.  While a large portion of the posterior probability is accessible to future multi-ton scale direct detection experiments, some of the region is below the lower limit from neutrino background, i.e. $10^{-48}$ cm$^2$~\cite{Baudis:2013qla,Newstead2013, Strigari:2009bq, Gutlein:2010tq}.  In contrast, given the more widely distributed posterior distribution of $\sigma_{\rm{SD}}$ and the lower sensitivity of experiments, a larger portion of the posterior is out of reach.

\subsubsection{Majorana Fermion}

\begin{figure}[htb]
\centering \hspace{9.20mm}
\includegraphics[width=8.07cm,trim=0mm 6.2mm -33mm 0mm,clip=true]{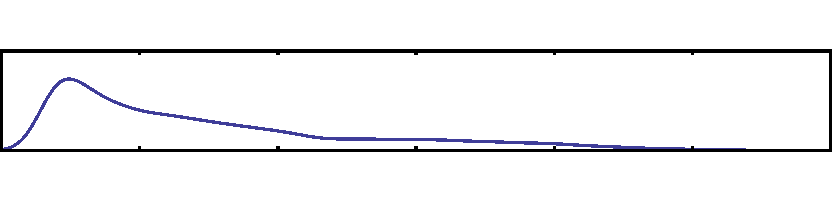} \\
\vspace{-3.9mm}
\includegraphics[width=8cm]{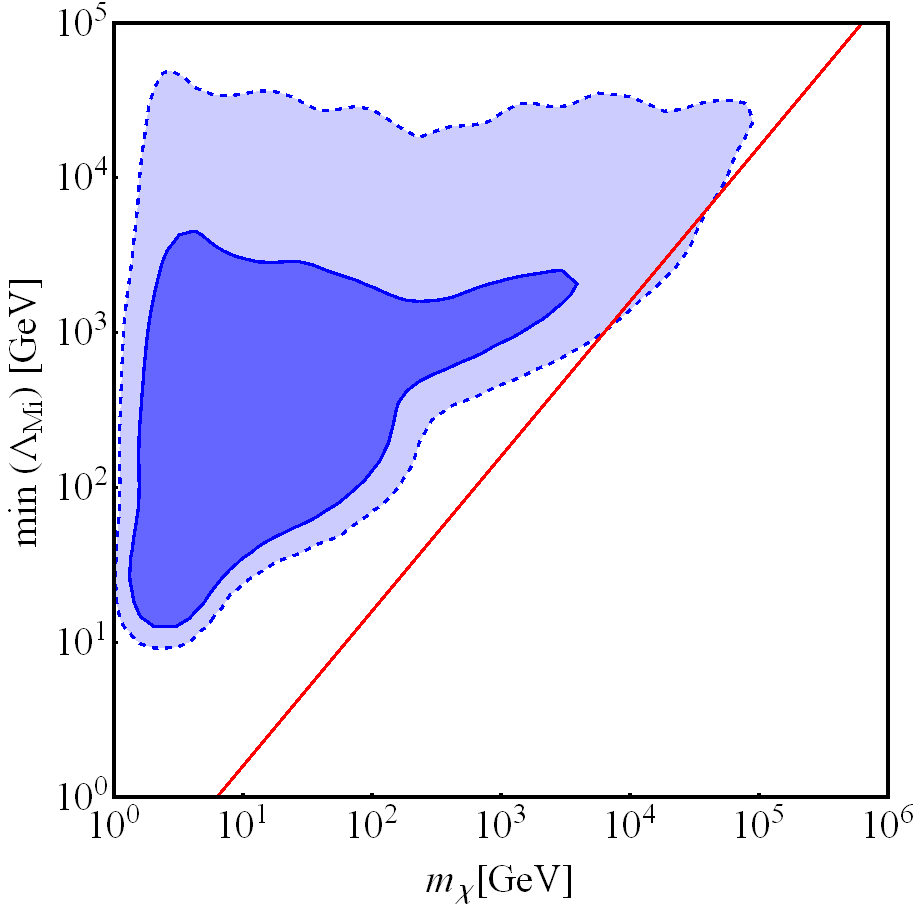} \hspace{-4.6mm}
\includegraphics[width=7.11cm,trim=0mm 10mm -8.mm 0mm,clip=true,angle=-90,origin=r]{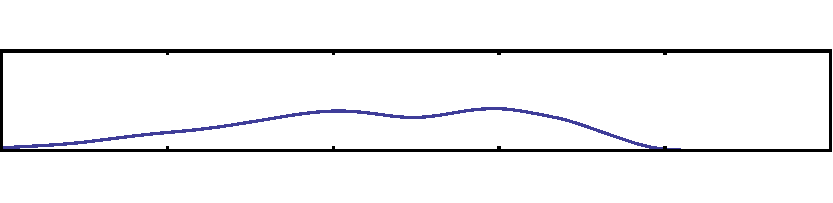}
\caption{Posterior probability distribution marginalized to the minimal $\Lambda$ scale and the mass of the Majorana fermion dark matter particle.  The light and dark regions correspond to 1 and 2$\sigma$ credible regions, respectively.  The red line corresponds to $\Lambda = \frac{m_\chi}{2\pi}$.}
\label{figLambdamMF}
\end{figure}

The posterior probability distribution obtained for the MF dark matter model, marginalized to the minimal cutoff scale vs. $m_\chi$ is shown in Fig.~\ref{figLambdamMF}.  Similarly to the DF case light dark matter mass favoured with $m_\chi$ below $10$ GeV, but with a more significant tail ranging to hundreds of GeV.  This result is fairly consistent with the expectations of a natural supersymmetric neutralino mass.  In the 2$\sigma$ credible region, the MF distribution stretches higher in mass ($\sim 80$ TeV) than in any other model, reflecting the results presented in Table~\ref{tableUpperLimits}.  The distribution marginalized to the mass is similar to that obtained for the DF model and due to the similarity of the Dirac and Majorana operators.

\begin{figure}[tb]
\centering
\includegraphics[scale=1,width=8cm]{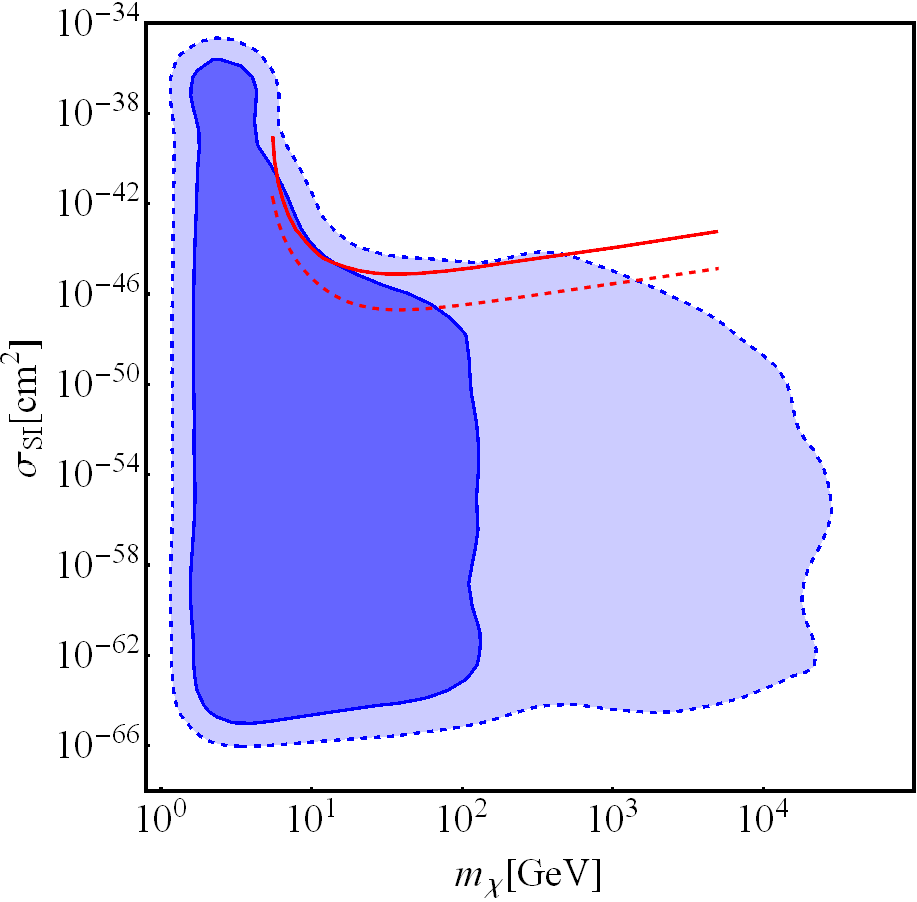}
\includegraphics[scale=1,width=8cm]{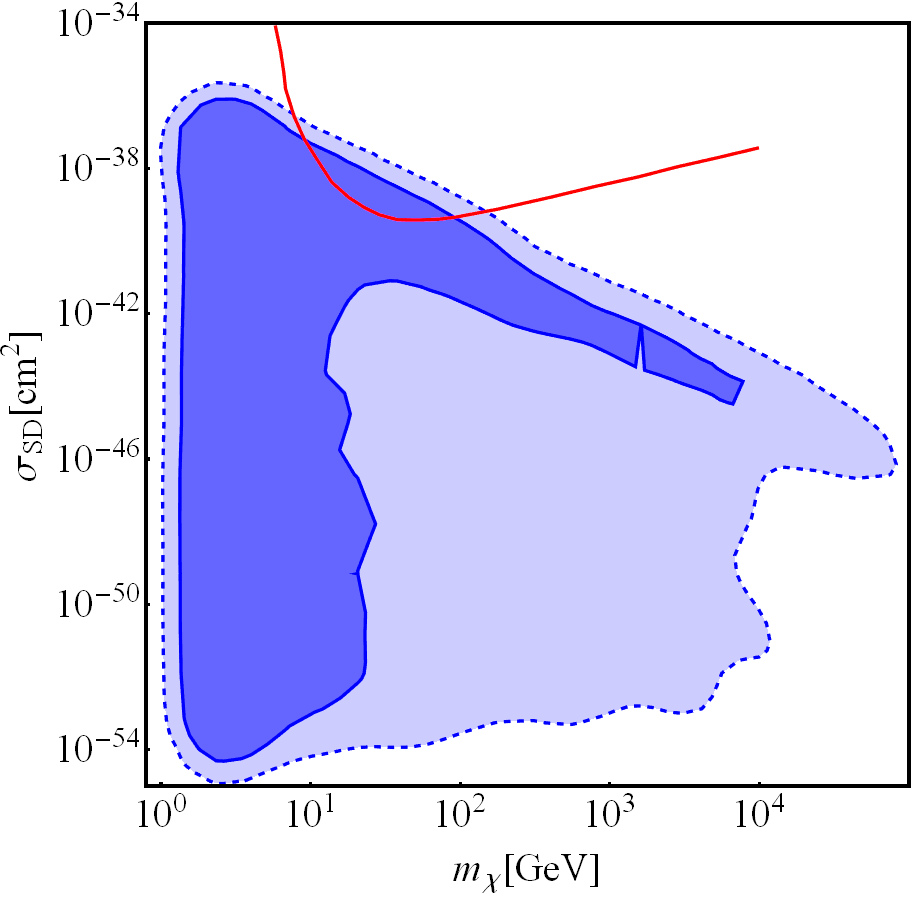}
\caption{Dark matter-proton elastic scattering cross section versus the mass of the Majorana fermion DM particle.  The light and dark regions correspond to 1 and 2$\sigma$ credible regions, respectively.  LUX (SI) and Xenon100 (SD) 90\% exclusion curves are shown in solid red, projected Xenon1t (SI, 2.6 tonne-years) limits are in dashed red.}
\label{figSigmamMF}
\end{figure}

The posterior distribution for the MF dark matter, marginalized to the $\sigma_{SI}$ ($\sigma_{SD}$) vs. $m_\chi$ plane, is shown in the left (right) panel of Fig.~\ref{figSigmamMF}.  Similar to the DF model, there are some high probability regions that will be probed at future experiments, but there remains a significant portion of the probability density far beyond the feasible reach of such experiments.  Comparison of the posterior distribution accessible to future experiments shows that spin dependent experiments appear to have more high probability region accessible to them.

\subsubsection{Complex Scalar}

\begin{figure}[htb]
\centering \hspace{9.20mm}
\includegraphics[width=8.07cm,trim=0mm 6.2mm -33mm 0mm,clip=true]{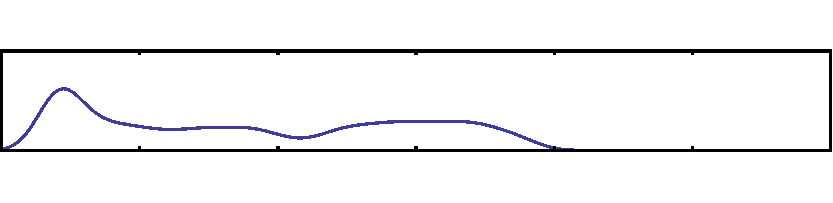} \\
\vspace{-3.9mm}
\includegraphics[width=8cm]{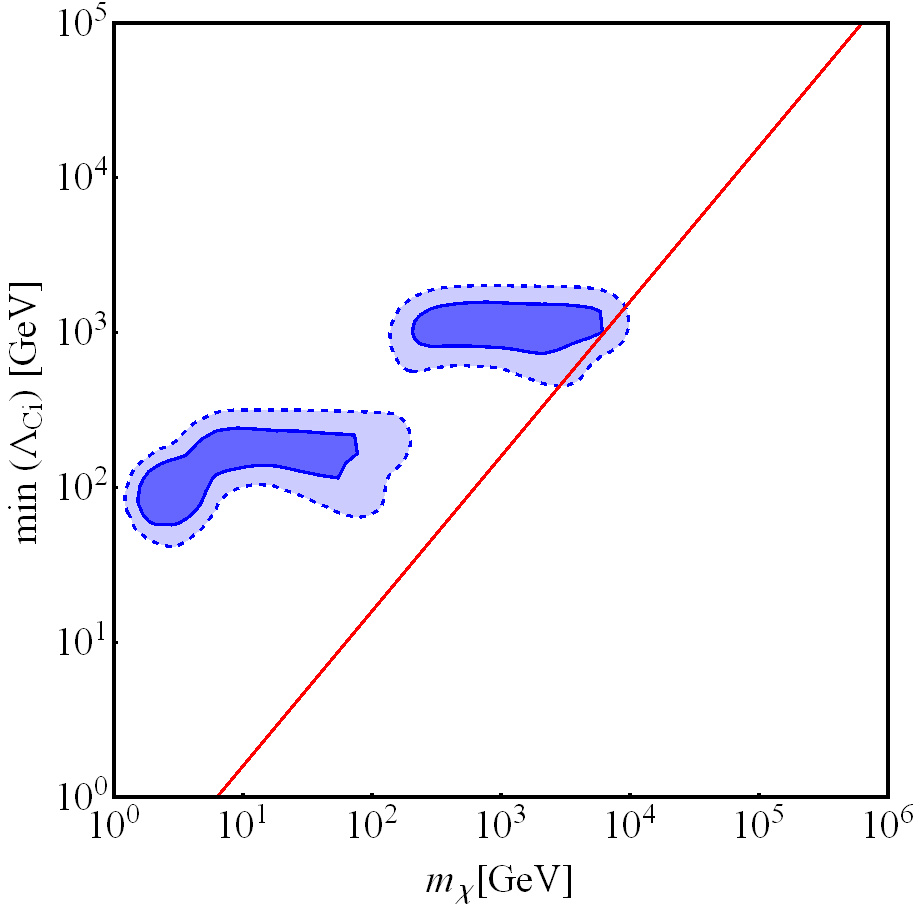} \hspace{-4.6mm}
\includegraphics[width=7.11cm,trim=0mm 10mm -8.mm 0mm,clip=true,angle=-90,origin=r]{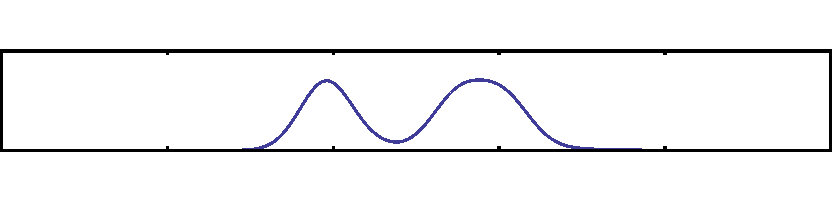}
\caption{Posterior probability distribution marginalized to the minimal $\Lambda$ scale and the mass of the complex scalar dark matter particle.  The light and dark regions correspond to 1 and 2$\sigma$ credible regions, respectively. The red line corresponds to $\Lambda = \frac{m_\chi}{2\pi}$.}
\label{figLambdamCS}
\end{figure}

The marginalized posterior distribution for the CS dark matter model is given in Fig.~\ref{figLambdamCS}.  The $1\sigma$ credible region in this case is bimodal and is much narrowed compared to the fermionic models, giving a preferred dark matter mass below 100 GeV or 200 GeV$<m_\chi<6.3$ TeV and the minimal cut-off is around 100 GeV or 1 TeV.

\begin{figure}[tb]
\centering
\includegraphics[scale=1,width=8cm]{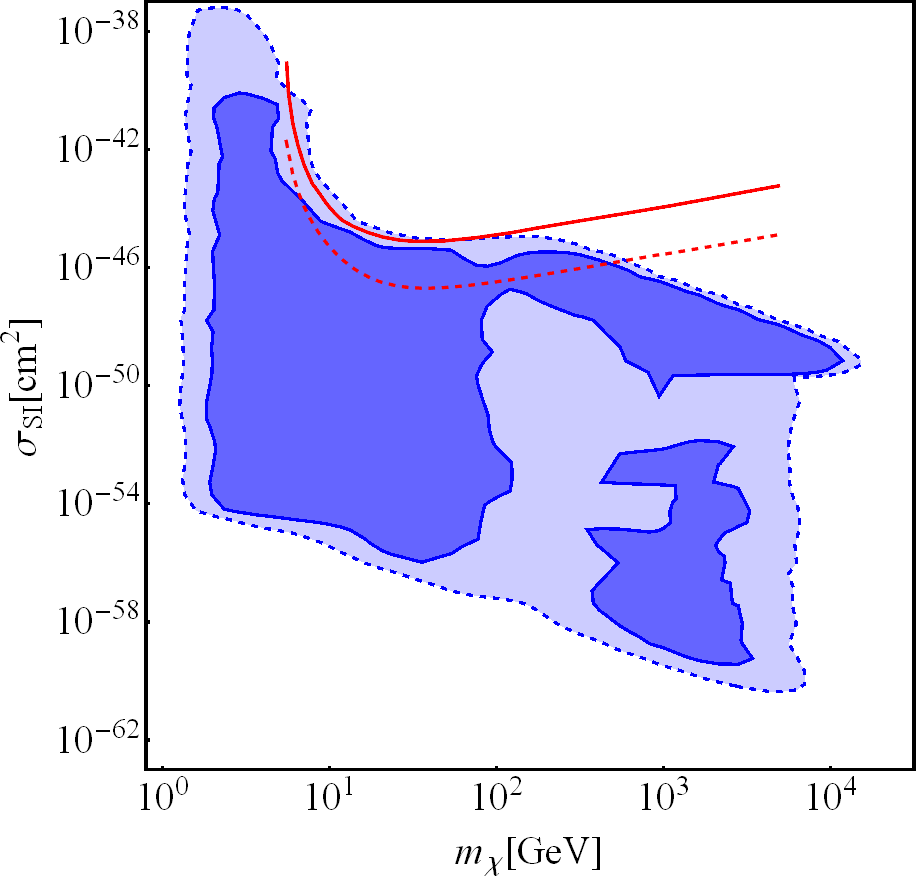}
\caption{Dark matter-proton elastic scattering cross section versus the mass of the complex scalar DM particle. The light and dark regions correspond to 1 and 2$\sigma$ credible regions, respectively. LUX (SI) 90\% exclusion curve is shown in solid red, projected Xenon1t (SI, 2.6 tonne-years) limits are in dashed red.}
\label{figSigmamCS}
\end{figure}

The posterior probability distribution marginalized to the $\sigma_{SI}$ vs. $m_\chi$ plane is shown in Fig.~\ref{figSigmamCS}.  The bimodal distribution is still evident, showing that the isolated higher-mass region is not discoverable in upcoming direct detection experiments.  Most of the $1\sigma$ region is out of the sensitivity range of future experiments, but unlike for fermionic dark matter, there is still large portion of high mass region that could be accessible for SD detection.

\subsubsection{Real Scalar}

\begin{figure}[htb]
\centering \hspace{9.20mm}
\includegraphics[width=8.07cm,trim=0mm 6.2mm -33mm 0mm,clip=true]{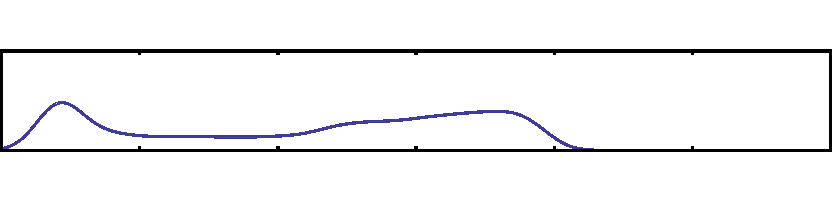} \\
\vspace{-3.9mm}
\includegraphics[width=8cm]{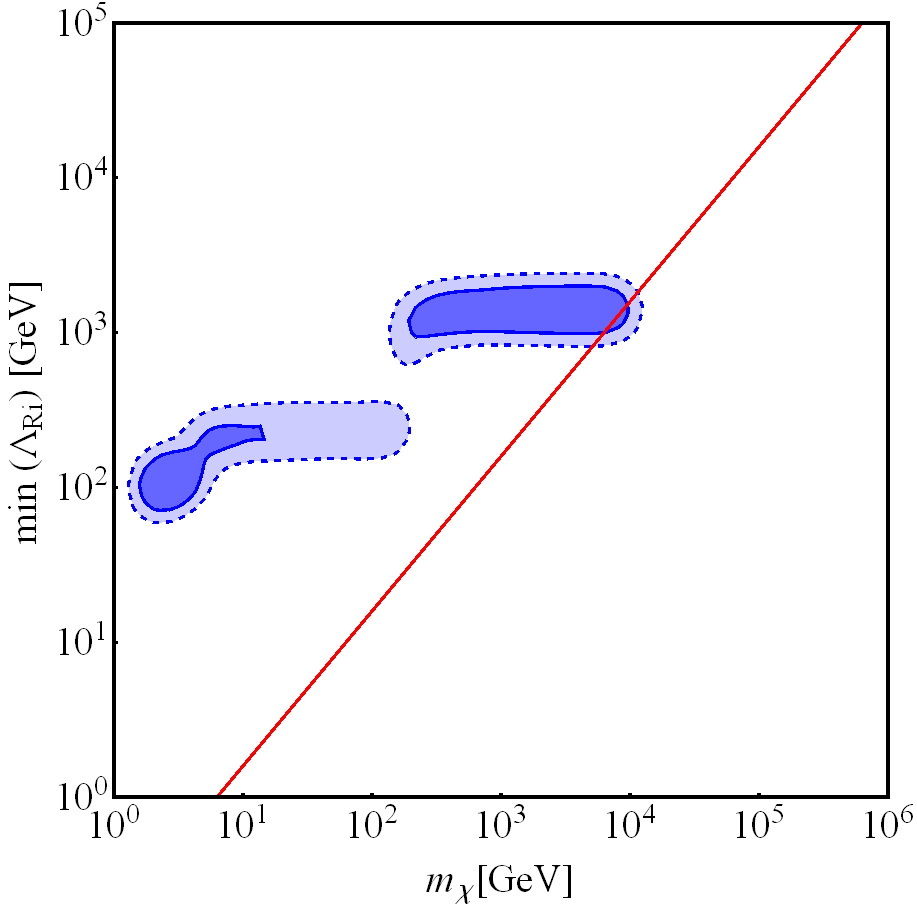} \hspace{-4.6mm}
\includegraphics[width=7.11cm,trim=0mm 10mm -8.mm 0mm,clip=true,angle=-90,origin=r]{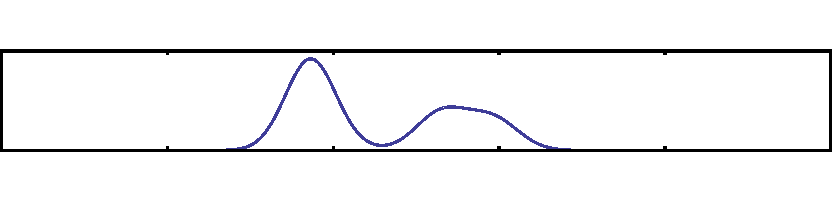}
\caption{Posterior probability distribution marginalized to the minimal $\Lambda$ scale and the mass of the real scalar dark matter particle. The light and dark regions correspond to 1 and 2$\sigma$ credible regions, respectively. The red line corresponds to $\Lambda = \frac{m_\chi}{2\pi}$.}
\label{figLambdamRS}
\end{figure}

The posterior probability distribution for the RS dark matter model, marginalized to the minimum cut-off scale, is shown in Fig.~\ref{figLambdamRS}.  The credible regions are almost identical to those of the CS dark matter.  The distribution is also bimodal in the 1$\sigma$ region with the mass $<10$ GeV and 200 GeV-10 TeV. The minimal cut-off scale is also strongly bimodal, with peaks at 150 GeV and 1.5 TeV and the higher region being more favoured.

\begin{figure}[tb]
\centering
\includegraphics[scale=1,width=8cm]{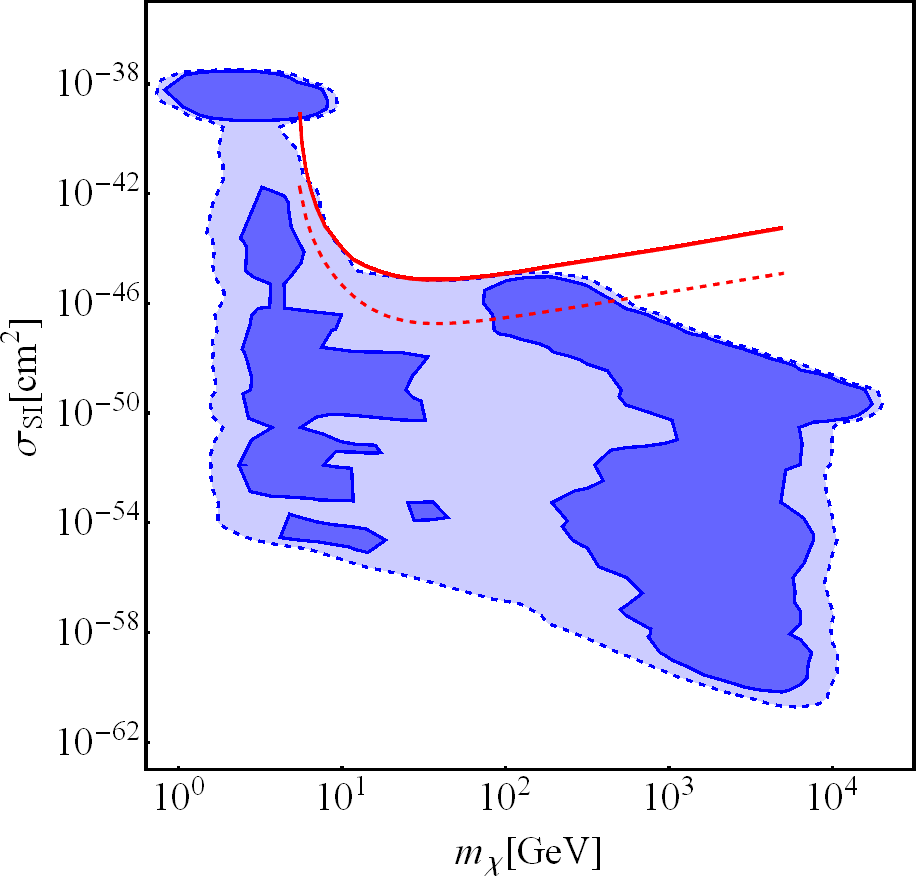}
\caption{Dark matter-proton elastic scattering cross section versus the mass of the real scalar dark matter particle.  The light and dark regions correspond to 1 and 2$\sigma$ credible regions, respectively. LUX (SI) 90\% exclusion curve is shown in solid red, projected Xenon1t (SI, 2.6 tonne-years) limits are in dashed red.}
\label{figSigmamRS}
\end{figure}

The posterior, marginalized to the $\sigma_{SI}$ vs. $m_\chi$ plane, is shown in Fig.~\ref{figSigmamRS}.  The distribution now shows three clearly favoured regions at 1$\sigma$, each of which can be partially probed at future experiments. The prospect of finding real scalar dark matter at future experiments is not optimistic, as only a very small amount of the posterior mass could be accessible.

\subsubsection{Vector Boson}

\begin{figure}[htb]
\centering \hspace{9.20mm}
\includegraphics[width=8.07cm,trim=0mm 6.2mm -33mm 0mm,clip=true]{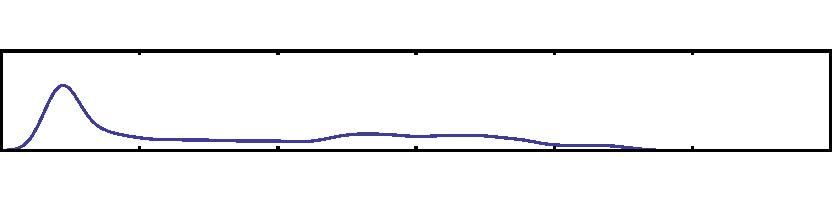} \\
\vspace{-3.9mm}
\includegraphics[width=8cm]{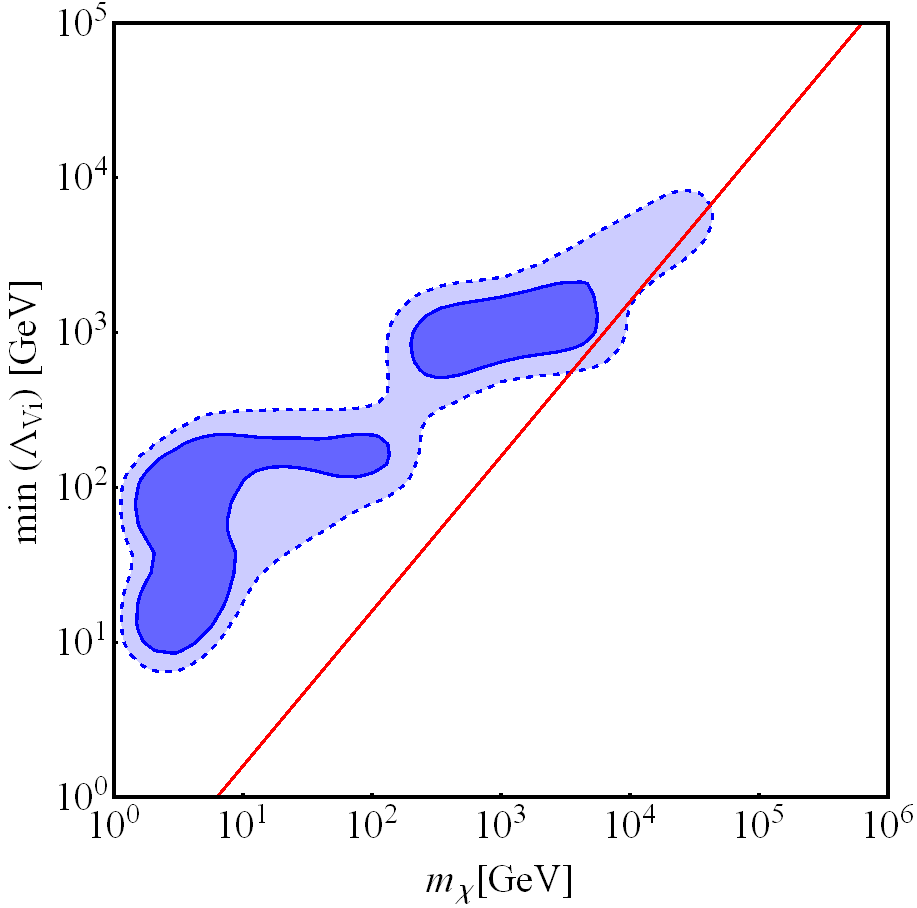} \hspace{-4.6mm}
\includegraphics[width=7.11cm,trim=0mm 10mm -8.mm 0mm,clip=true,angle=-90,origin=r]{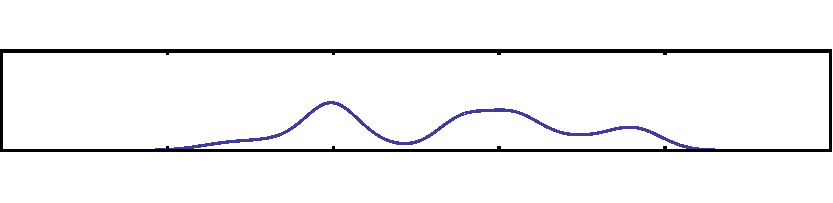}
\caption{Posterior probability distribution marginalized to the minimal $\Lambda$ scale and the mass of the vector boson dark matter particle.  The light and dark regions correspond to 1 and 2$\sigma$ credible regions, respectively. The red line corresponds to $\Lambda = \frac{m_\chi}{2\pi}$.}
\label{figLambdamVB}
\end{figure}

Fig.~\ref{figLambdamVB} shows the resulting marginalized posterior probability distribution for the VB dark matter, in the minimum cut-off vs. $m_\chi$ plane.  While the distribution is bimodal, when marginalized to the mass parameter, it is clear that the low mass, under $10$ GeV, is the favoured region.  At 2$\sigma$, however, the region extends to 40 TeV.  The minimum cut-off scale is almost trimodal, but is mostly favoured to be around 100 GeV or 1 TeV.

\begin{figure}[tb]
\centering
\includegraphics[scale=1,width=8cm]{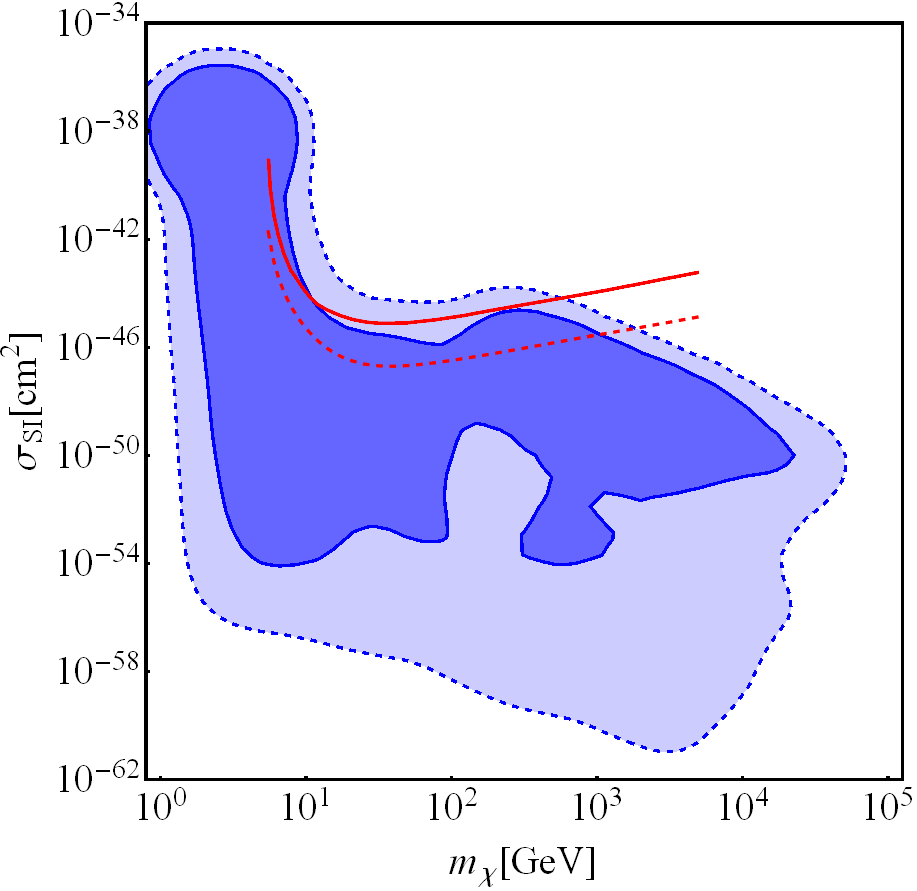}
\caption{Dark matter-proton elastic scattering cross section versus the mass of the vector boson DM particle. The light and dark regions correspond to 1 and 2$\sigma$ credible regions, respectively. LUX (SI) 90\% exclusion curve is shown in solid red, projected Xenon1t (SI, 2.6 tonne-years) limits are in dashed red.}
\label{figSigmamVB}
\end{figure}

The posterior, marginalized to the $\sigma_{SI}$ vs. $m_\chi$ plane, is shown in Fig.~\ref{figSigmamVB}.  With a large amount of the posterior mass lying within the sensitivity range of future experiments, VB dark matter is the most accessible one among all the models considered here.

\subsection{Astrophysical and Collider Experiments}

Besides relic abundance and direct detection experiments, astrophysical and collider experiments give additional constrains on the DM models.  We find, however, that astrophysical experiments have negligible effect beyond the constraints discussed above.  This is not surprising, as it is known that typically large boost factors are expected to see an astrophysical signal of dark matter annihilation \cite{Ackermann:2011wa, Hektor:2012kc, Yuan:2013eja, Garny:2013ama}.  For this reason we only include indirect detection constraints in our analysis in the case of Dirac fermion dark matter for illustrative purposes.\footnote{The proper implementation of indirect detection limits in the effective field theory context is quite complicated as seen from Refs.~\cite{Jin:2013nta,Fedderke:2013pbc,Alves:2014yha}.}

Collider experiments are not much more constraining than the relic density and direct detection together~\cite{tsai_dark_2013}, except in the low mass region.  However, there are two relevant operators for DF (D5 and D8) where collider limits are competitive and readily available~\cite{Aad:2013oja}.
There are caveats when applying LHC constraints in the effective field theory framework~\cite{Dreiner:2013vla, Busoni:2013lha, Busoni:2014sya}. For illustrative purposes Fig.~\ref{figSigmamDFcomp} below shows the effect of adding collider (D5 and D8) and astrophysical constraints (Fermi-LAT~\cite{Ackermann:2012rg}) to a Dirac fermion scan.

\begin{figure}[tb]
\centering
\includegraphics[scale=1,width=8cm]{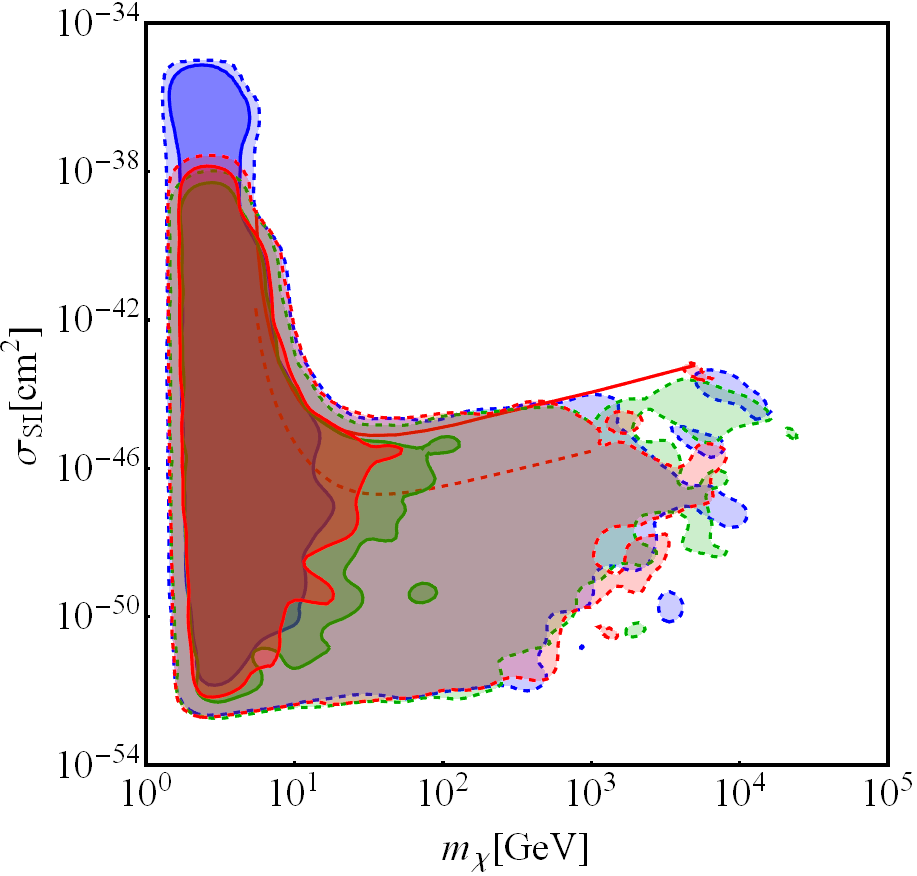}
\includegraphics[scale=1,width=8cm]{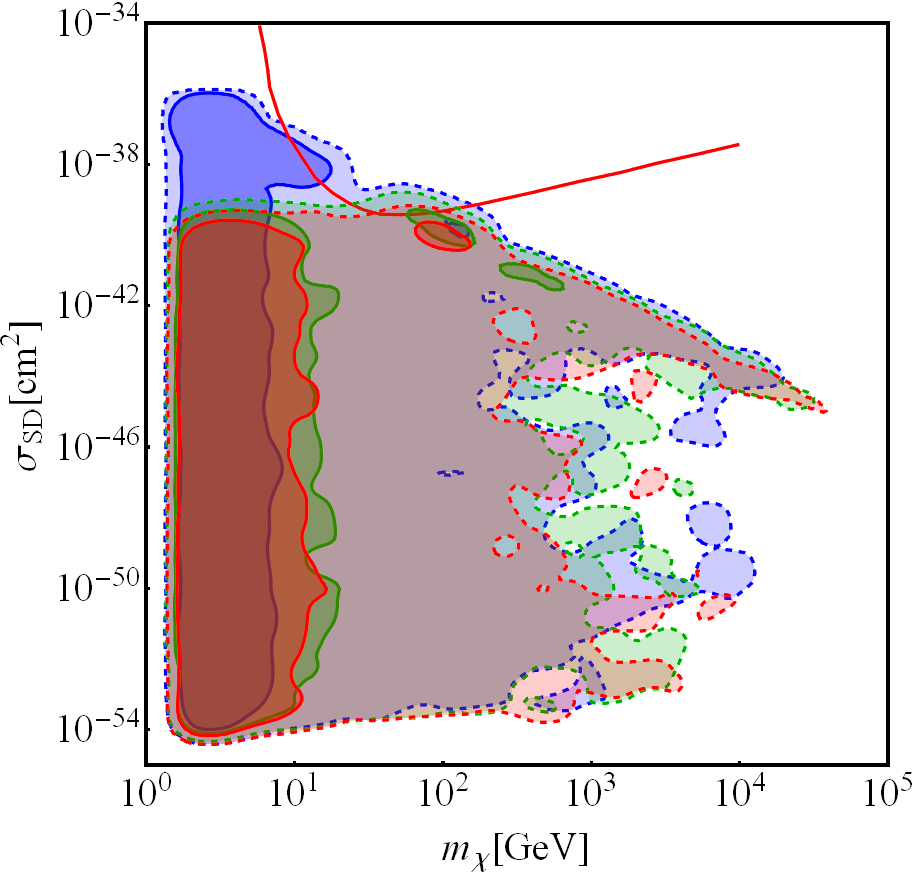}
\caption{Dark matter-proton elastic scattering cross section versus the mass of the Dirac fermion dark matter particle.  The spin independent cross section is shown in the left frame, and the spin dependent cross section in the right frame.  Blue regions have direct detection bounds only (as above), green regions have direct detection bounds and collider bounds, while red shows direct, indirect and collider bounds. The light and dark regions correspond to 1 and 2$\sigma$ credible regions, respectively.}
\label{figSigmamDFcomp}
\end{figure}

\section{Summary}
\label{Sum}

We carried out a comprehensive and model-independent analysis for various dark matter candidates based on the effective field theory framework.  By using the Bayesian inference we calculated posterior probabilities for the parameters of the five different effective dark matter models.  To extract these probabilities we relied on multiple observations, such as the dark matter abundance and direct detection limits.  Inclusion of further experimental constraints can make our analysis even more informative.

Some general conclusions observed in all the explored models are
\begin{itemize}
\item The scale where new physics cuts off the effective theory cannot be indefinitely large as a consequence of preventing overproducing dark matter relic abundance.  Their universal upper limit is at the level of $10^3-10^4$ GeV at 1$\sigma$ CL.
\item A light dark matter mass is favoured, in the region of 10-100 GeV, which agrees with expectations of naturalness of new physics \cite{Giudice:2013yca, Vecchi:2013iza}, and various putative dark matter signals \cite{Daylan:2014rsa}.
\item While future direct detection and collider searches will be able to probe or constrain these models further, considerable part of the feasible parameter space is out of their reach.  While this might not give an optimistic outlook for discovering dark matter in the near future, it forces us to consider new avenues for the experimental and theoretical exploration of the dark matter problem.
\item With the exception of the vector boson model, the most favoured operators in all the models contain the $\bar f \gamma^\mu f $ bilinear.  This suggests a skewed sense of parity between the Standard Model and the dark sector.  The emergence of such generic features shows the power of the effective field theory approach combined with Bayesian inference to solve the dark matter problem.
\end{itemize}

\acknowledgments
The work of C.B.~and T.L.~was supported by the ARC Centre of Excellence for Particle Physics at the Terascale. J.N.~was supported by an Australian Postgraduate Award and the U.S.~Department of Energy.


\appendix

\section{Function $F_{x_F}(x)$ and the total annihilation cross section}

The function $F_{x_F}(x)$ is defined as, and approximated via a power expansion of the Bessel functions around $x=0$:
\begin{eqnarray}
F_{x_F}(x) & = & \sqrt{\frac{x\pi}{2}}\int^{1/x_F}_0\frac{K_1(\frac{x}{y})}{y K_2(\frac{1}{y})^2}dy \nonumber \\
& \approx &\frac{x\pi}{2}\sqrt{\frac{\pi}{x-2}} \left(1-\mathrm{Erf}\left[\sqrt{(x-2)x_F}\right]\right)+x\pi\left(\frac{3}{8 x}-\frac{15}{4}\right) \nonumber \\
& & \left(e^{(2-x)x_F} \sqrt{x_F^{-1}}-\sqrt{x-2} \sqrt{\pi } \left(1-\mathrm{Erf}\left[\sqrt{(x-2)x_F}\right]\right)\right) \nonumber \\
& & +\frac{x\pi}{3} \left(\frac{285}{32}-\frac{45}{32 x}-\frac{15}{128 x^2}\right)   \nonumber \\
& & \left(2 \sqrt{\pi } \left(1-\mathrm{Erf}\left[\sqrt{(x-2)x_F}\right]\right) (x-2)^{3/2} \right. \nonumber \\
& & \left. +e^{(2-x)x_F} \left(x_F^{-3/2}+2 (2-x) \sqrt{x_F^{-1}}\right)\right). \nonumber \\
\end{eqnarray}

The total cross sections of dark matter annihilating to a pair of SM fermions, given the operators in Tables~\ref{tableDFoperators},\ref{tableMFoperators},\ref{tableVBoperators} and \ref{tableRCSoperators}, are

\begin{eqnarray}
\sigma_{\rm ann}^{DF}&=&\sum_{f=l,q}{N_C \over 48\pi s\Lambda_{D1}^6\Lambda_{D2}^6\Lambda_{D3}^6\Lambda_{D4}^6\Lambda_{D5}^4\Lambda_{D6}^4\Lambda_{D7}^4\Lambda_{D8}^4}\sqrt{s-4m_f^2\over s-4m_\chi^2}\nonumber \\
&&[\Lambda_{D1}^6(\Lambda_{D2}^6(\Lambda_{D3}^6(4\Lambda_{D4}^6(\Lambda_{D5}^4(
\Lambda_{D6}^4(\Lambda_{D7}^4(4m_f^2(7m_\chi^2-s)+s(s-4m_\chi^2))\nonumber \\
&&+\Lambda_{D8}^4(4m_f^2-s)(2m_\chi^2+s))
+\Lambda_{D7}^4\Lambda_{D8}^4(2m_f^2+s)(4m_\chi^2-s))\nonumber \\
&&+\Lambda_{D6}^4\Lambda_{D7}^4\Lambda_{D8}^4(2m_f^2+s)(2m_\chi^2+s))
-24\Lambda_{D4}^3\Lambda_{D5}^4\Lambda_{D6}^4\Lambda_{D7}^4\Lambda_{D8}^2m_f^2m_\chi s\nonumber \\
&&+3\Lambda_{D5}^4\Lambda_{D6}^4
\Lambda_{D7}^4\Lambda_{D8}^4m_f^2s^2)+3\Lambda_{D4}^6\Lambda_{D5}^4\Lambda_{D6}^4\Lambda_{D7}^4\Lambda_{D8}^4
m_f^2s(s-4m_\chi^2))\nonumber \\
&&+3\Lambda_{D3}^6\Lambda_{D4}^6\Lambda_{D5}^4\Lambda_{D6}^4\Lambda_{D7}^4\Lambda_{D8}^4m_f^2s(s-4m_f^2))\nonumber \\
&&-3\Lambda_{D2}^6\Lambda_{D3}^6\Lambda_{D4}^6\Lambda_{D5}^4\Lambda_{D6}^4\Lambda_{D7}^4\Lambda_{D8}^4
(4m_f^2-s)(s-4m_\chi^2)],
\end{eqnarray}
\begin{eqnarray}
\sigma_{\rm ann}^{MF}&=&\sum_{f=l,q}{N_C \over 48\pi s\Lambda_{M1}^6\Lambda_{M2}^6\Lambda_{M3}^6\Lambda_{M4}^6\Lambda_{M5}^4\Lambda_{M6}^4}\sqrt{s-4m_f^2\over s-4m_\chi^2}\nonumber \\
&&[\Lambda_{M1}^6(\Lambda_{M2}^6(\Lambda_{M3}^6(4\Lambda_{M4}^6(\Lambda_{M5}^4(4m_f^2(7m_\chi^2-s)+s(s-4m_\chi^2))
-\Lambda_{M6}^4(2m_f^2+s)(4m_\chi^2-s))\nonumber \\
&&-3\Lambda_{M5}^4\Lambda_{M6}^4m_f^2s^2)+3\Lambda_{M4}^6\Lambda_{M5}^4
\Lambda_{M6}^4m_f^2s(s-4m_\chi^2))+3\Lambda_{M3}^6\Lambda_{M4}^6\Lambda_{M5}^4\Lambda_{M6}^4m_f^2s(s-4m_f^2))\nonumber \\
&&-3\Lambda_{M2}^6\Lambda_{M3}^6\Lambda_{M4}^6\Lambda_{M5}^4\Lambda_{M6}^4m_f^2(4m_f^2-s)(s-4m_\chi^2)],
\end{eqnarray}
\begin{eqnarray}
\sigma_{\rm ann}^{CS}&=&\sum_{f=l,q}{N_C \over 48\pi s\Lambda_{C1}^4\Lambda_{C2}^4\Lambda_{C3}^4\Lambda_{C4}^4}\sqrt{s-4m_f^2\over s-4m_\chi^2}\nonumber \\
&&[6\Lambda_{C2}^4\Lambda_{C3}^4\Lambda_{C4}^4m_f^2(s-4m_f^2)-\Lambda_{C1}^4\Lambda_{C2}^4\Lambda_{C3}^4(2m_f^2(8m_\chi^2-5s)+s(s-4m_\chi^2))\nonumber \\
&&+\Lambda_{C1}^4\Lambda_{C2}^4\Lambda_{C4}^4(2m_f^2+s)(4m_\chi^2-s)
-12\Lambda_{C1}^4\Lambda_{C2}^2\Lambda_{C3}^4\Lambda_{C4}^2m_f^2s
+6\Lambda_{C1}^4\Lambda_{C3}^4\Lambda_{C4}^4m_f^2s],
\end{eqnarray}
\begin{eqnarray}
\sigma_{\rm ann}^{RS}=\sum_{f=l,q}{N_C\over 8\pi s\Lambda_{R1}^4\Lambda_{R2}^4}\sqrt{s-4m_f^2\over s-4m_\chi^2}m_f^2(\Lambda_{R2}^4(s-4m_f^2)+s\Lambda_{R1}^4),
\end{eqnarray}
\begin{eqnarray}
\sigma_{\rm ann}^{VB}&=&\sum_{f=l,q}{N_C \over 144\pi s\Lambda_{V1}^4\Lambda_{V2}^4\Lambda_{V3}^8\Lambda_{V4}^8}\sqrt{s-4m_f^2\over s-4m_\chi^2}m_f^2\nonumber \\
&&[6\Lambda_{V1}^4\Lambda_{V2}^2\Lambda_{V3}^8\Lambda_{V4}^4s(2m_\chi^2-s)+8\Lambda_{V1}^4\Lambda_{V3}^8\Lambda_{V4}^8s
+\Lambda_{V2}^4(\Lambda_{V3}^8(\Lambda_{V1}^4s(6m_\chi^4-4m_\chi^2s+s^2)\nonumber \\
&&+8\Lambda_{V4}^8(s-4m_f^2))-\Lambda_{V1}^4
\Lambda_{V4}^8(4m_f^2-s)(6m_\chi^4-4m_\chi^2s+s^2)\nonumber \\
&&+6\Lambda_{V1}^2\Lambda_{V3}^4\Lambda_{V4}^8(4m_f^2-s)(s-2m_\chi^2))],
\end{eqnarray}
where $N_C=1 (3)$ for SM leptons (quarks).

\section{Bayesian Inference}

In this section we briefly summarize the statistical underpinnings of our analysis.  Given two non-exclusive propositions, $A$ and $B$, the plausibility of these two propositions in light of some prior information, $I$, is $P(A|I)$ and $P(B|I)$.  The plausibility that they are both correct is given by the conditional probability \begin{equation}
 P(AB|I) = P(A|BI)P(B|I) .
\end{equation}
The symmetry of the conditional probability under the exchange of $A$ and $B$ leads to Bayes' theorem:
\begin{equation}
 P(A|BI) = \frac{P(B|AI)P(A|I)}{P(B|I)} .
\end{equation}
We introduce the standard names used for these quantities in parameter extraction.  If $A$ represents a hypothesis then $P(A|I)$ is called the prior probability.  This represents the plausibility of our hypothesis given information prior the observation $B$.  The likelihood function $P(B|AI)$ represents how accurately the hypothesis can replicate the data.  The posterior probability $P(A|BI)$ quantifies the plausibility of the hypothesis $A$ given the data $B$.  The evidence $P(B|I)$ serves to normalize the posterior.

For the posterior to represent a proper probability distribution we must have a complete set of independent hypotheses such that
\begin{equation}
\sum_{i}P(H_i) = 1 ,
\label{eqnH}
\end{equation}
with $A = H_1$.
For theoretical models with a continuous parameter $\theta$ the above formula can be recast in the form
\begin{equation}
\mathcal{P}(\theta|B,I) = \frac{\mathcal{L}(B|\theta,I)\pi(\theta,I)}{\epsilon(B,I)} .
\label{eqnBayes}
\end{equation}
The posterior distribution in the latter form can be used to estimate the most likely parameter region of a theory.
In the case of a continuous parameter the evidence is calculated via an integral over the full parameter space
\begin{equation}
\epsilon(B,I) = \int_\theta \mathcal{L}(B|\theta,I)\pi(\theta,I) d\theta .
\end{equation}

Marginalization is performed by integrating the posterior over various parameters in the higher dimensional parameter space
\begin{equation}
\mathcal{P}(\theta_j) = \int \prod_{i\neq j} d\theta_i \mathcal{P}(\theta_i).
\end{equation}

\subsection{Likelihood Functions}

Whenever an experimental central value is available with an uncertainty, we cast the likelihood function in the form of a Gaussian distribution centered on the measured value with standard deviation equal to the uncertainty:
\begin{equation}
\mathcal{L}_i(d|\theta,I) =  \frac{1}{\sqrt{2\pi}\sigma}\mathrm{Exp}\left(-\frac{(x(\theta) - d)^2}{2\sigma^2}\right).
\end{equation}
For experiments that only place a bound on a particular parameter, the likelihood function will take the form of a complementary error function:
\begin{equation}
\mathcal{L}_i(d|\theta,I) =  \frac{1}{2}\mathrm{Erfc} \left(\frac{x(\theta) - d}{2\sigma}\right).
\end{equation}

The composite likelihood combines likelihood functions for various data points $d_i$ at the parameter point $\theta$
\begin{equation}
\mathcal{L}_{\mathrm{T}}(D|\theta,I) = \prod_i \mathcal{L}_i(d_i|\theta,I).
\end{equation}

We combine experimental and theoretical uncertainties in quadrature, and assume that theoretical calculations of relic density and direct detection have an error of 10\% throughout the whole parameter space.

\section{Posterior probability distributions for individual operator}
Here we include plots of the posterior probability distributions marginalized to the $\Lambda_i$ versus DM particle mass.

\begin{figure}[htb]
\centering
\includegraphics[width=5.4cm]{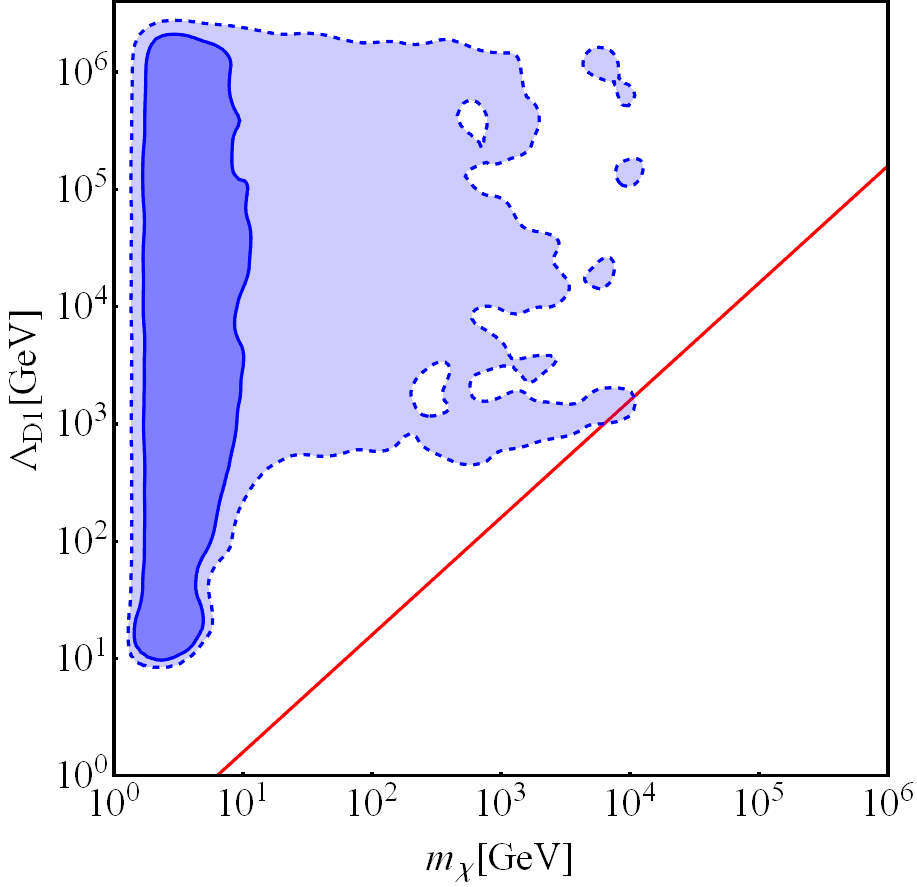}
\includegraphics[width=5.4cm]{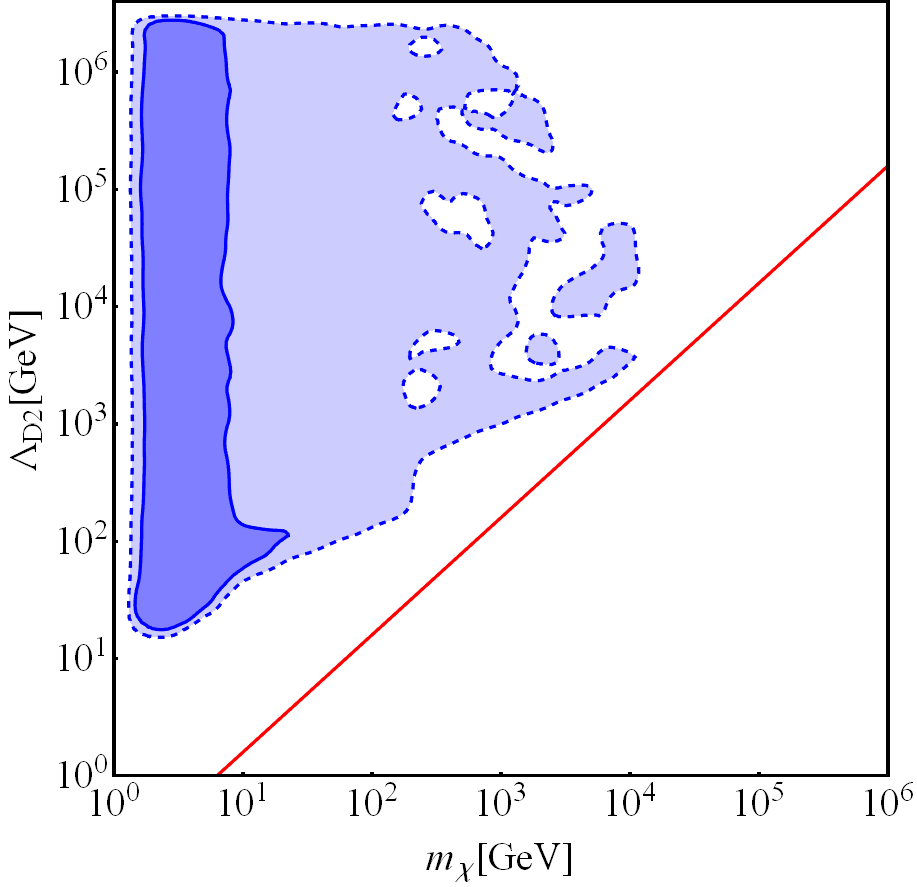}
\includegraphics[width=5.4cm]{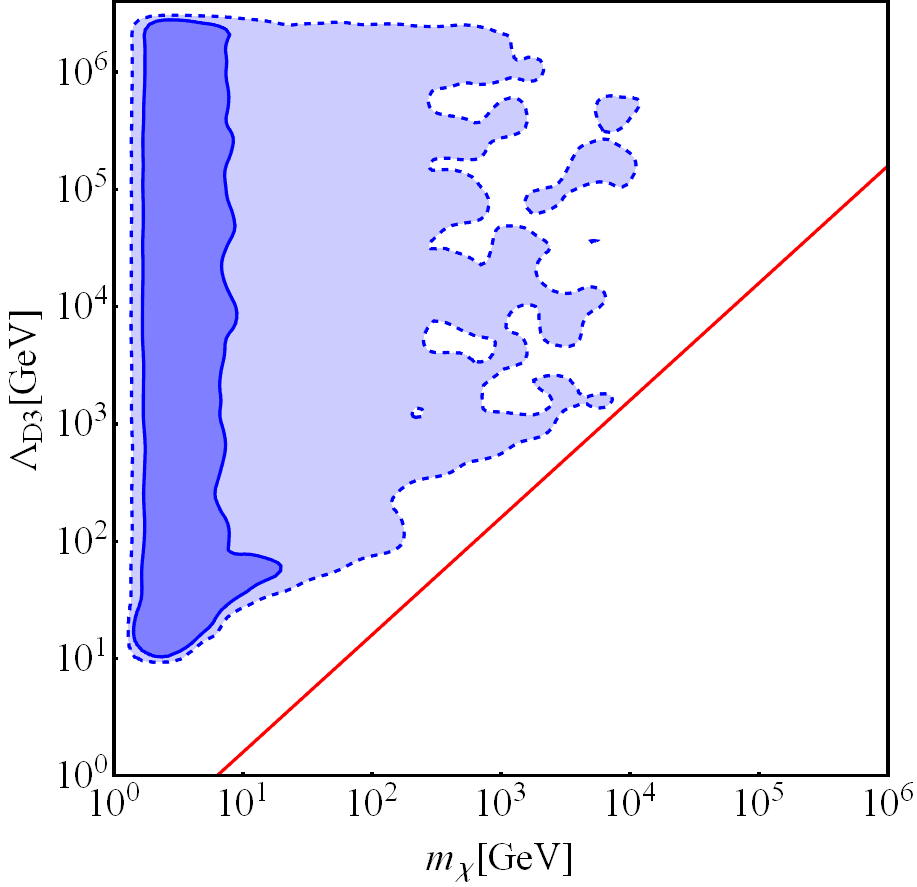}
\includegraphics[width=5.4cm]{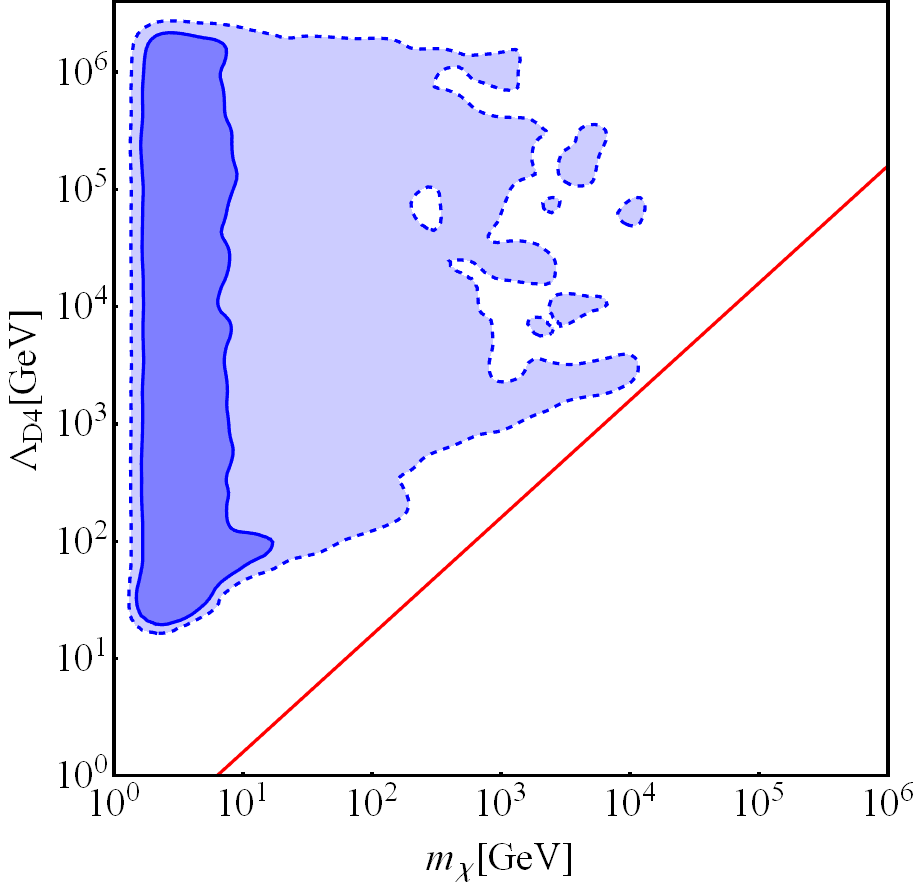}
\includegraphics[width=5.4cm]{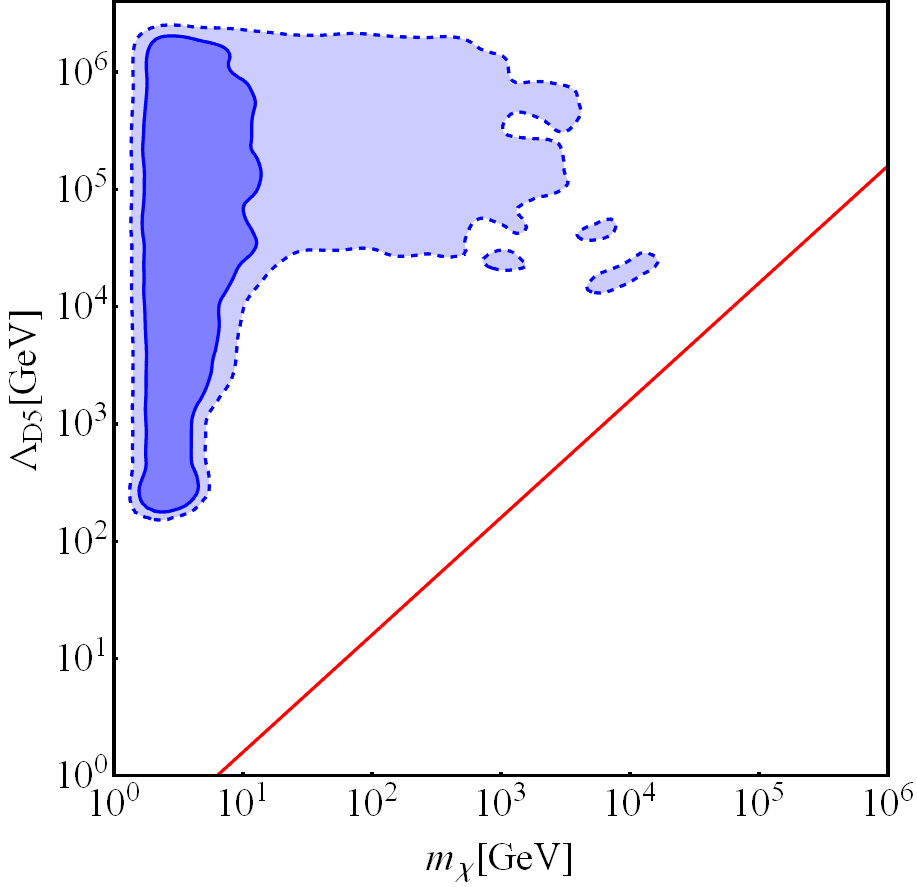}
\includegraphics[width=5.4cm]{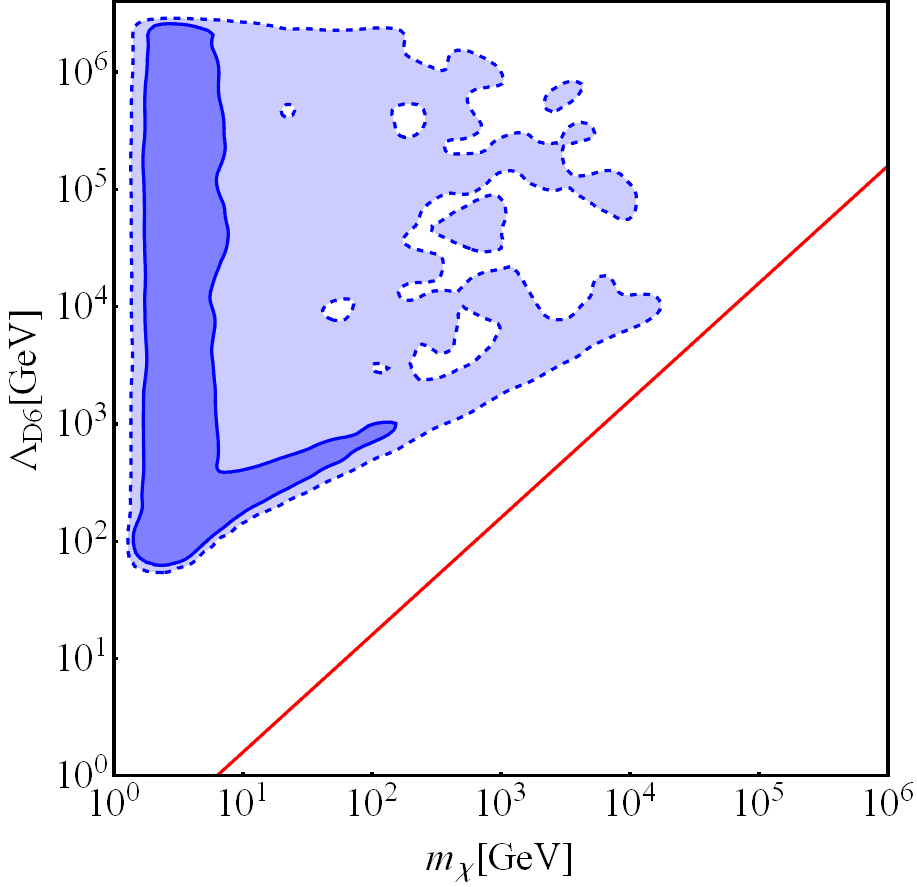}
\includegraphics[width=5.4cm]{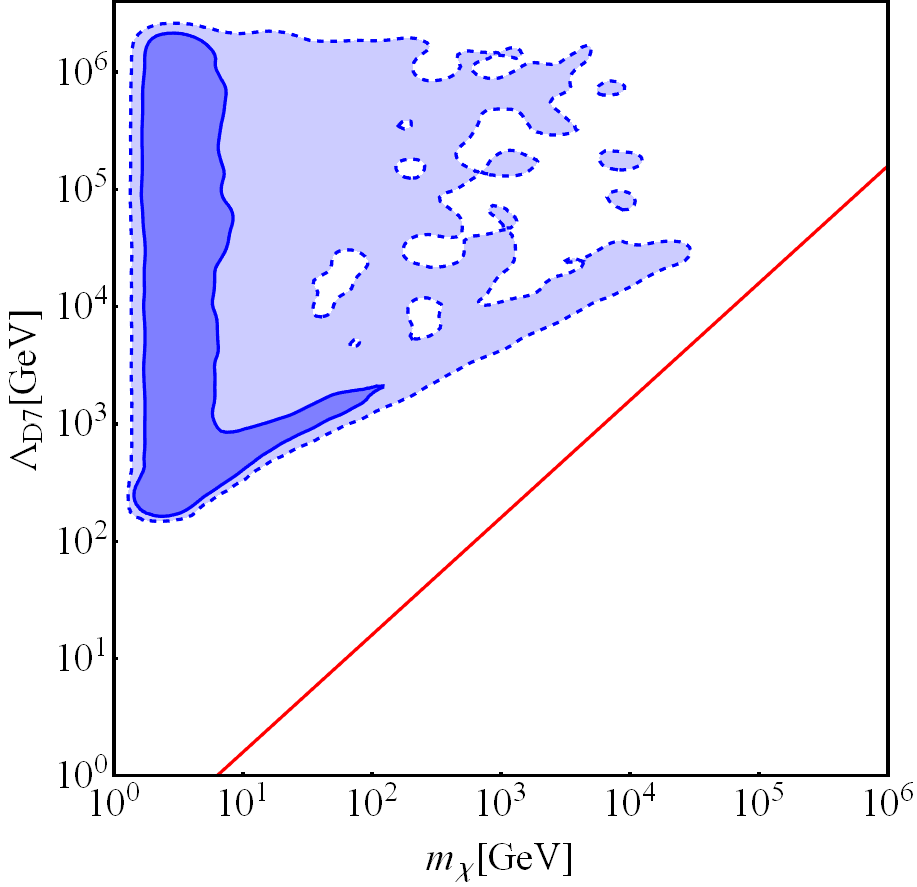}
\includegraphics[width=5.4cm]{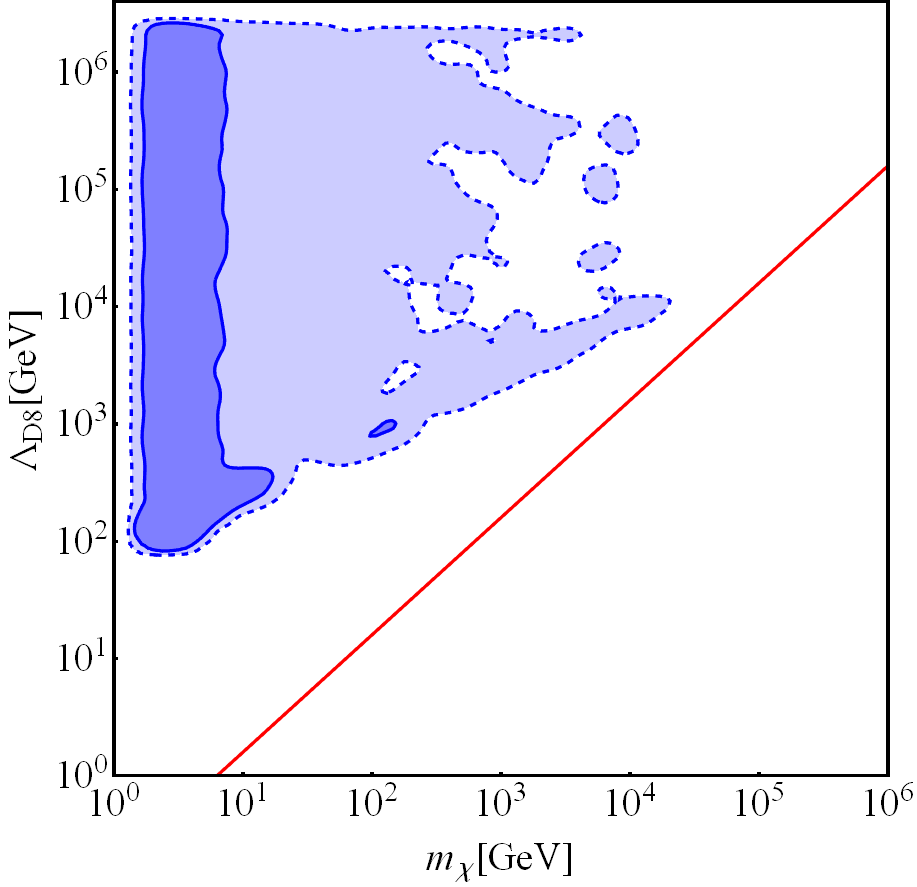}
\caption{Posterior probability distribution marginalized to the $\Lambda_{Di}$ scale ($i=1-8$) and the mass of the Dirac fermion DM particle.}
\label{figLambdamDFall}
\end{figure}

\begin{figure}[htb]
\centering
\includegraphics[width=5.4cm]{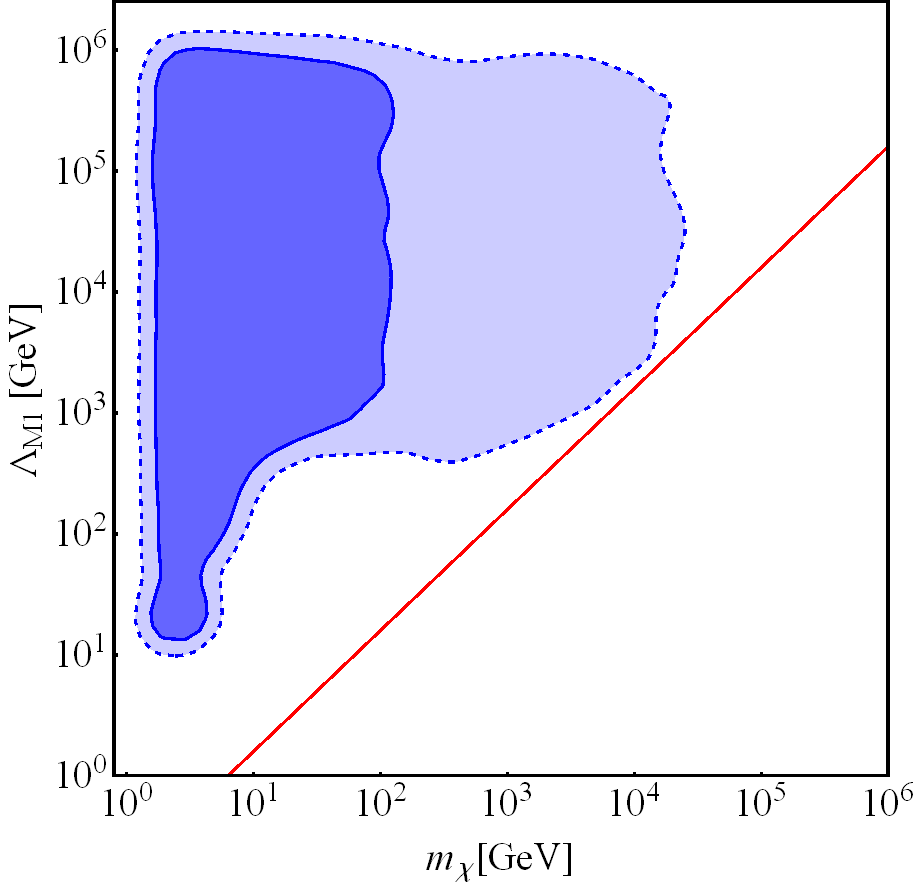}
\includegraphics[width=5.4cm]{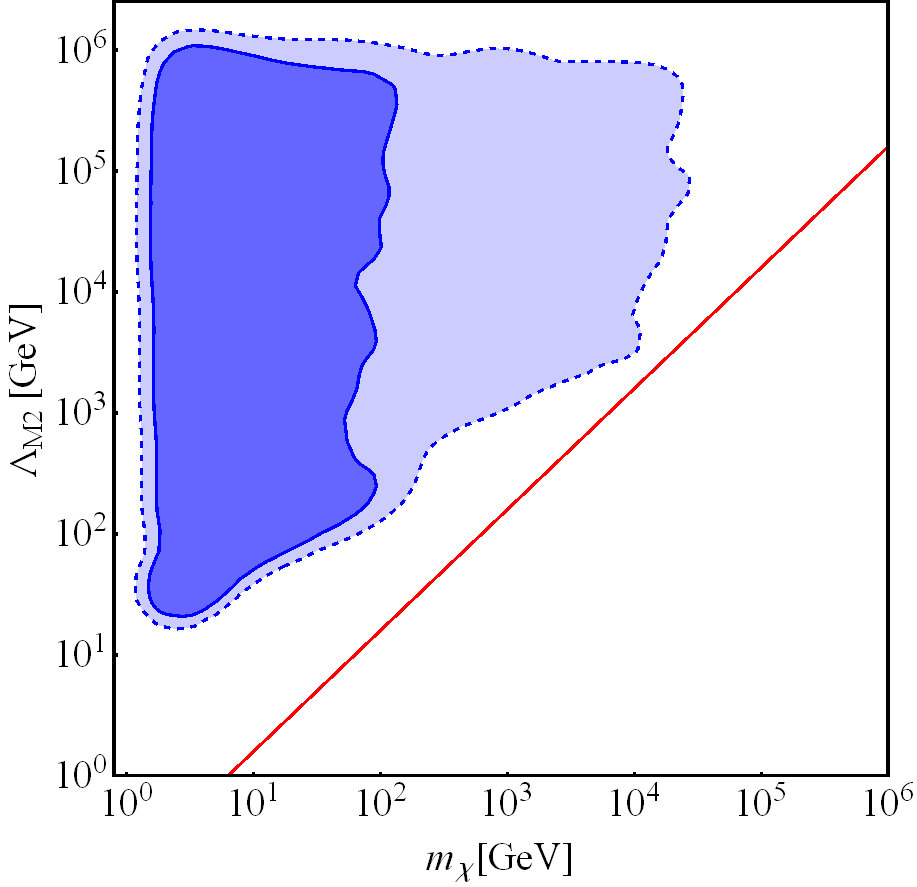}
\includegraphics[width=5.4cm]{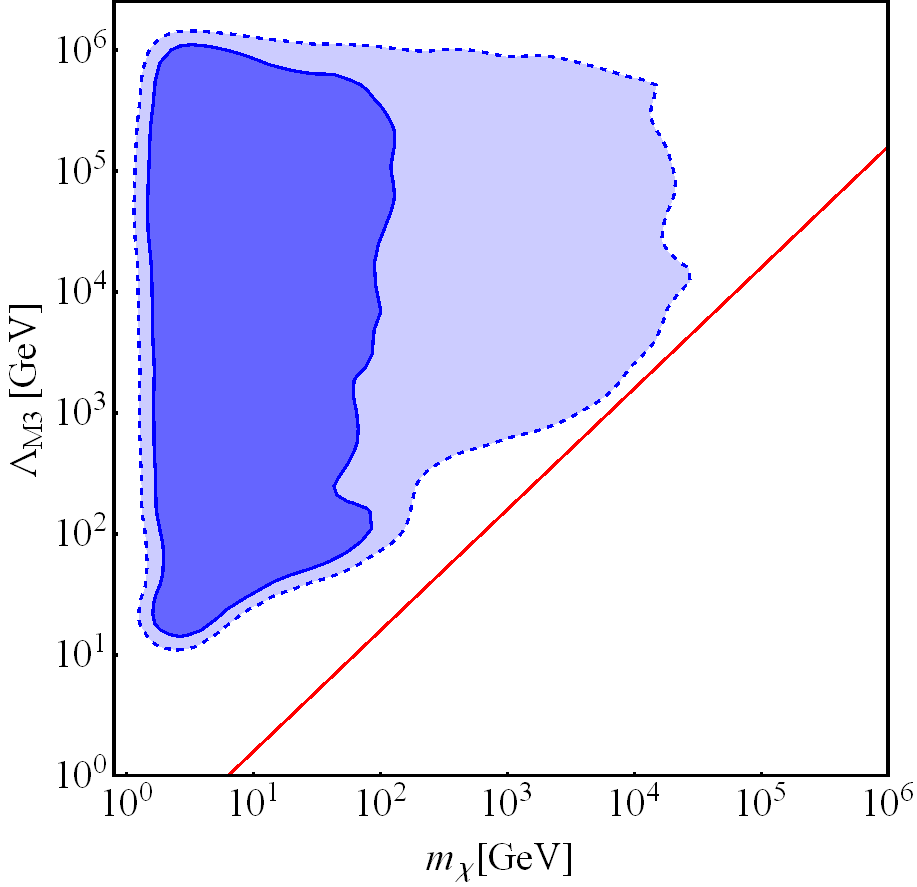}
\includegraphics[width=5.4cm]{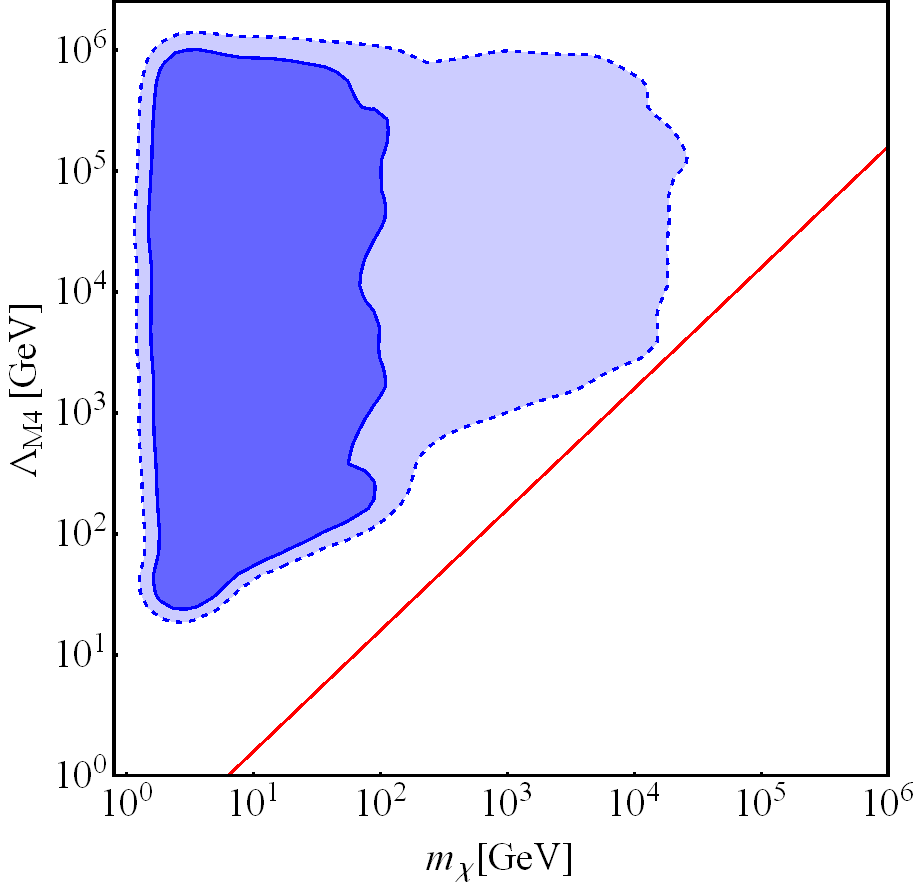}
\includegraphics[width=5.4cm]{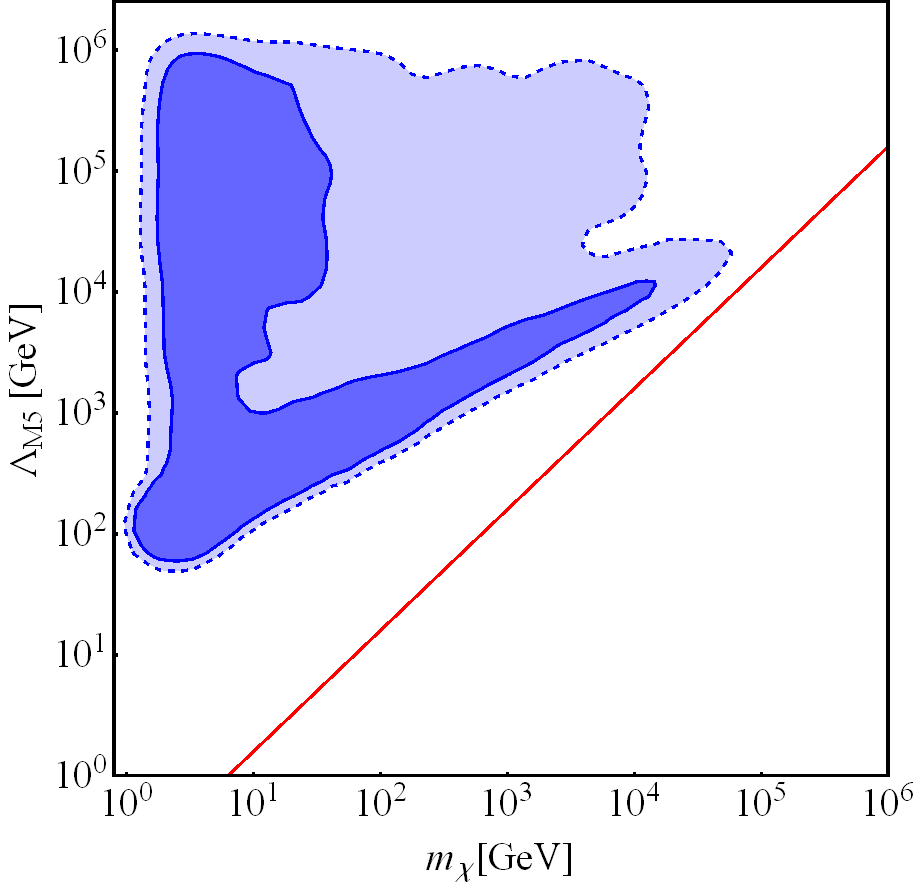}
\includegraphics[width=5.4cm]{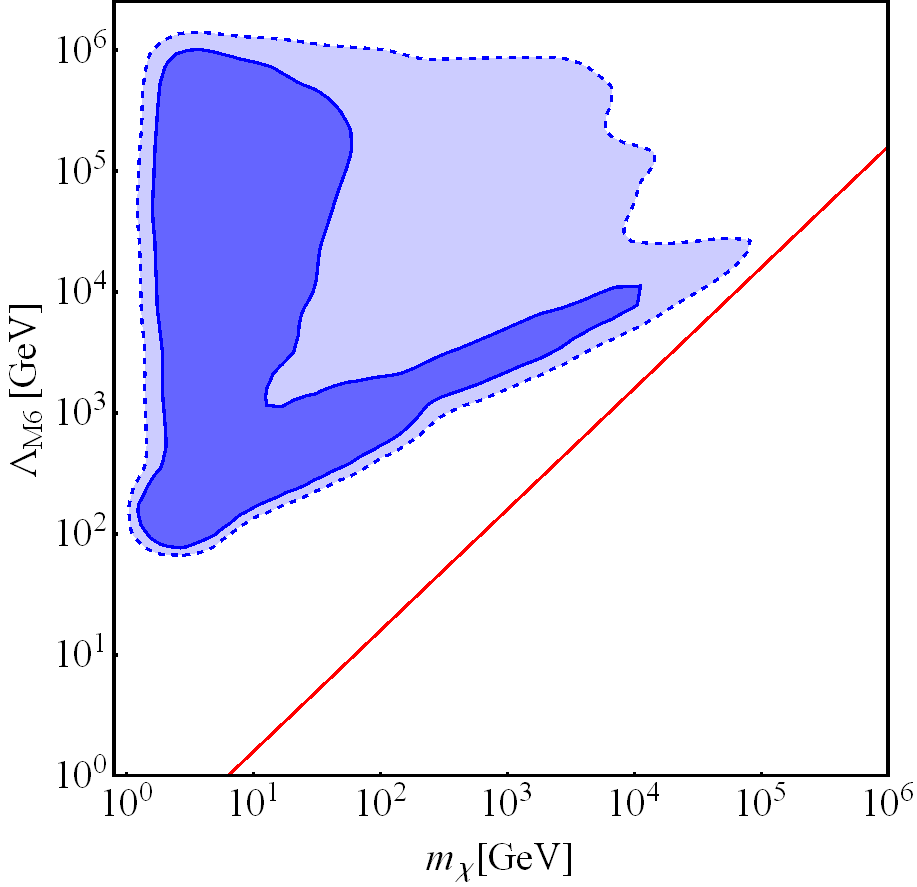}
\caption{Posterior probability distribution marginalized to the $\Lambda_{Mi}$ scale ($i=1-6$) and the mass of the Majorana fermion DM particle.}
\label{figLambdamMFall}
\end{figure}

\begin{figure}[htb]
\centering
\mbox{\includegraphics[width=5.4cm]{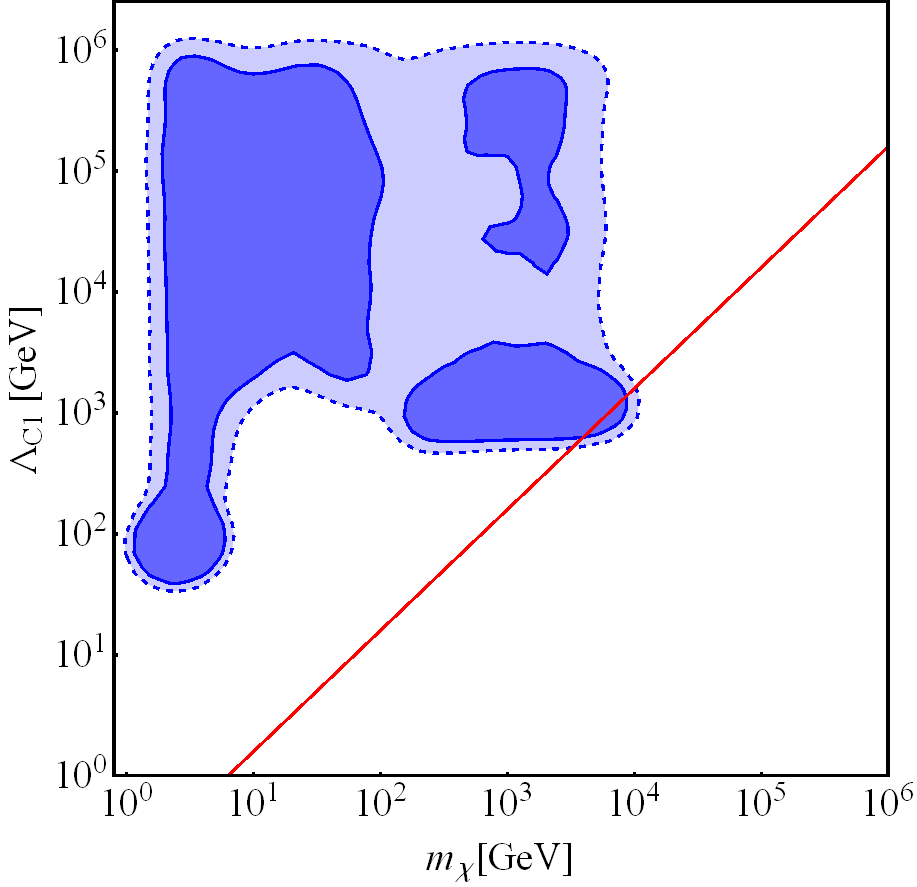},
\includegraphics[width=5.4cm]{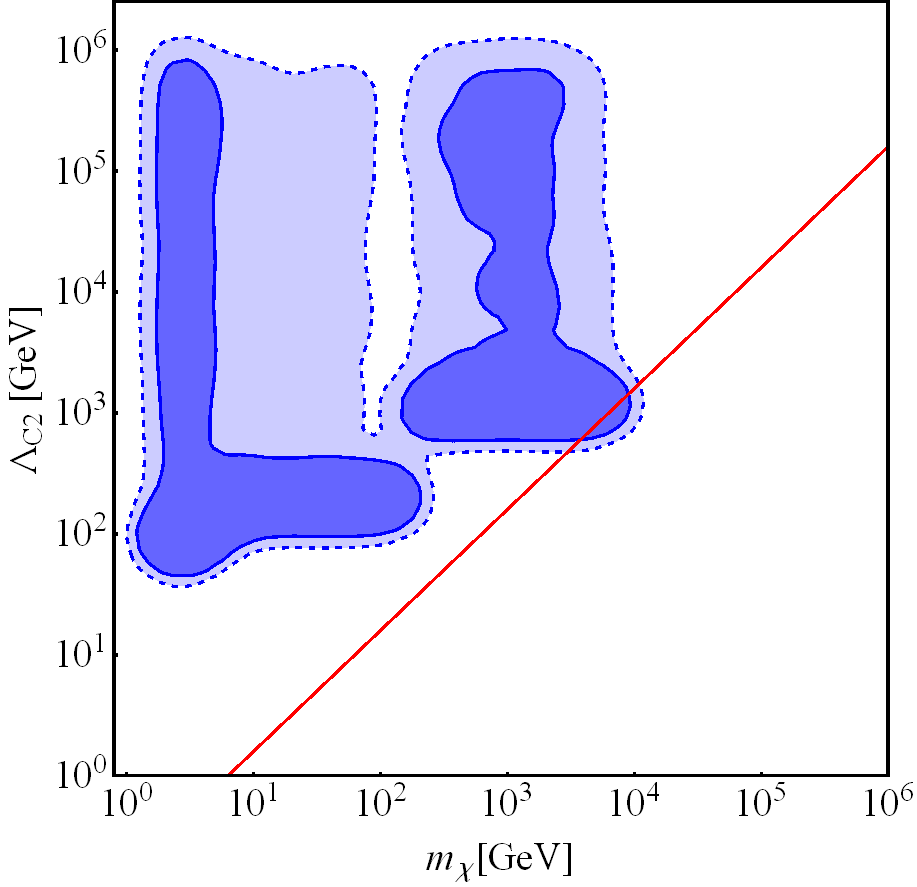}}
\includegraphics[width=5.4cm]{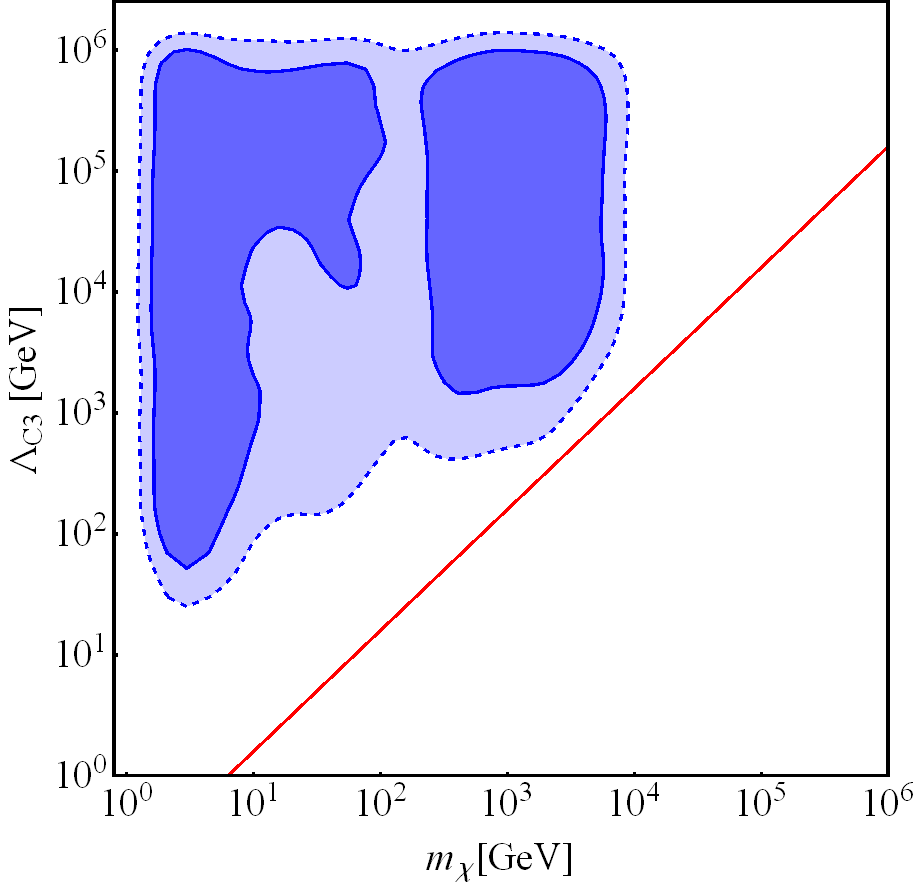}
\includegraphics[width=5.4cm]{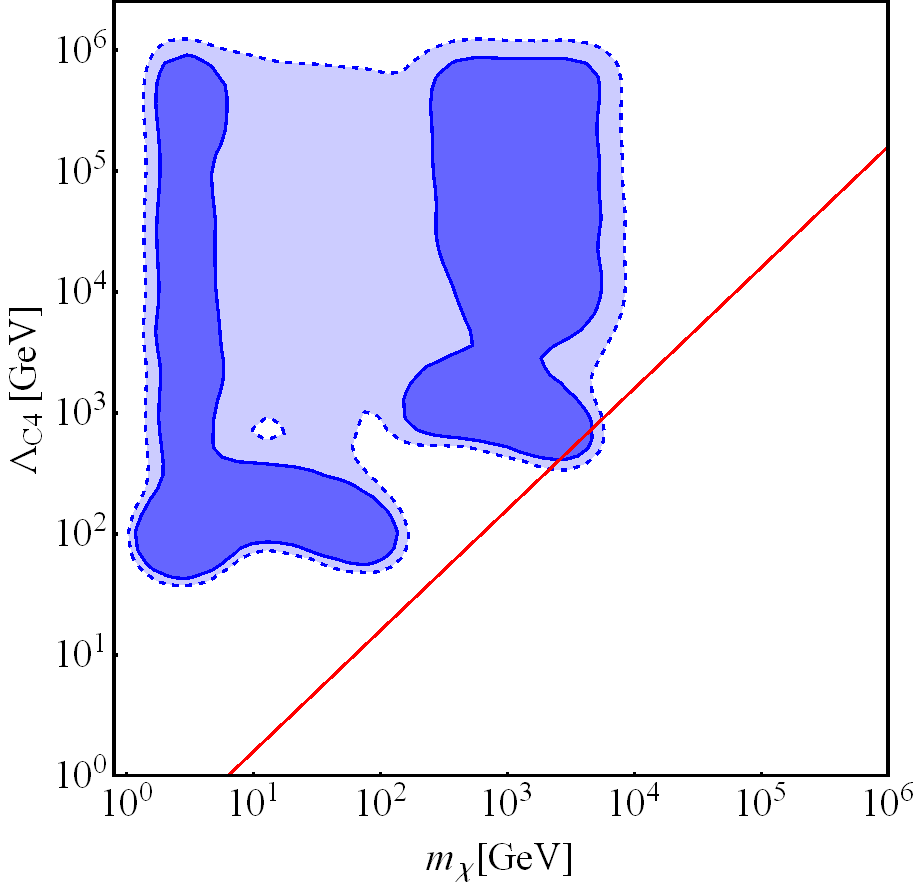}
\caption{Posterior probability distribution marginalized to the $\Lambda_{Ci}$ scale ($i=1-4$) and the mass of the complex scalar DM particle.}
\label{figLambdamCSall}
\end{figure}

\begin{figure}[htb]
\centering
\includegraphics[width=5.4cm]{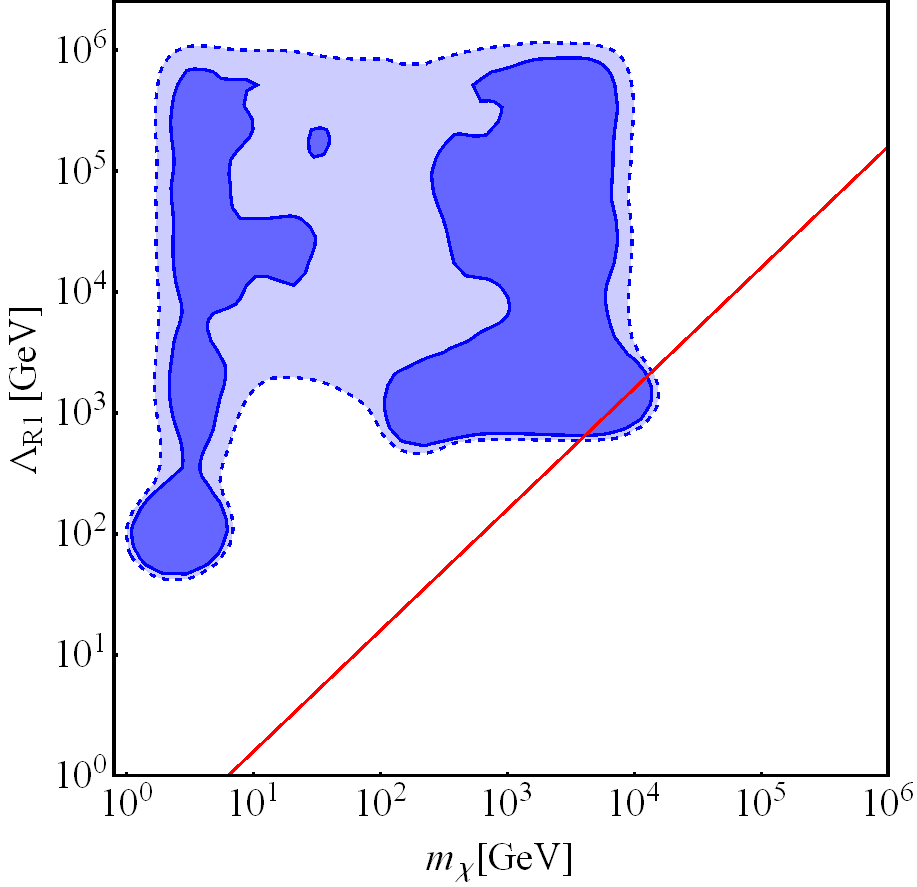}
\includegraphics[width=5.4cm]{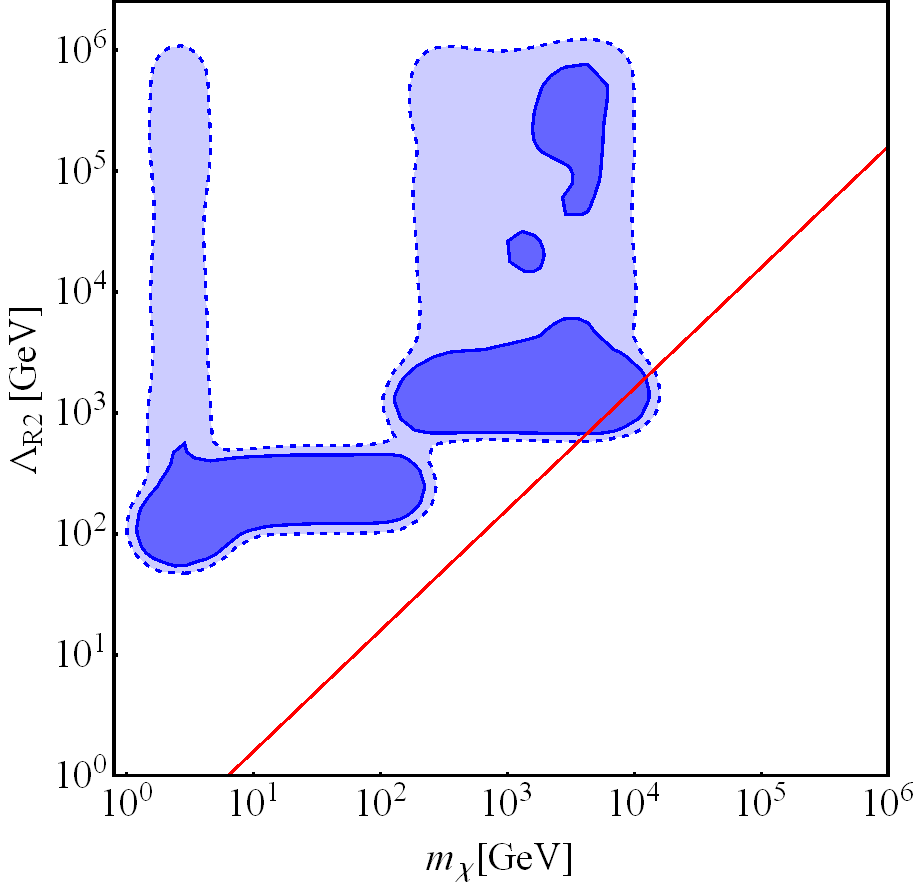}
\caption{Posterior probability distribution marginalized to the $\Lambda_{Ri}$ scale ($i=1-2$) and the mass of the real scalar DM particle.}
\label{figLambdamRSall}
\end{figure}

\begin{figure}[htb]
\centering
\mbox{\includegraphics[width=5.4cm]{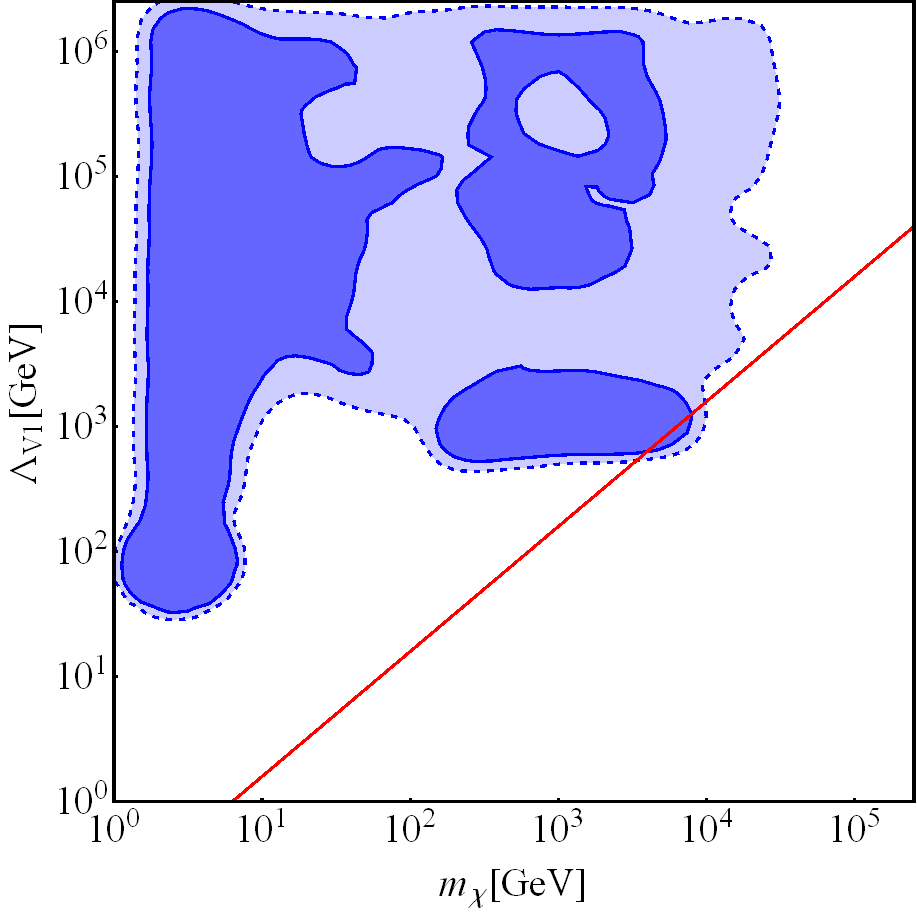},
\includegraphics[width=5.4cm]{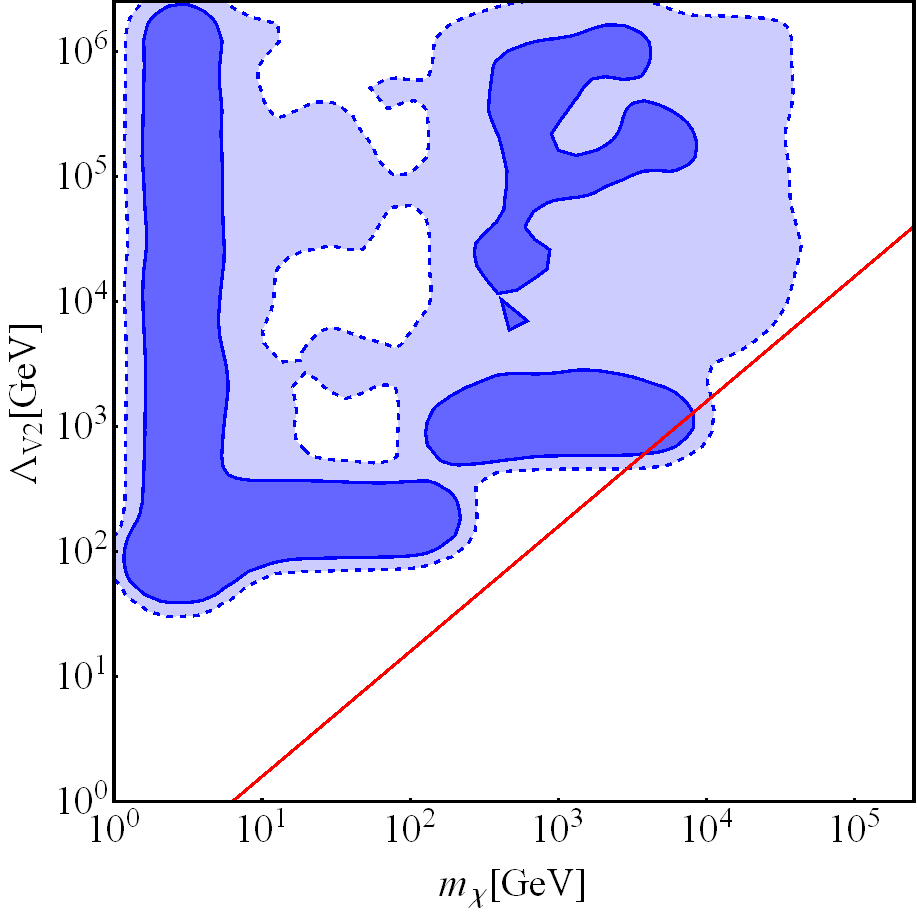}}
\includegraphics[width=5.4cm]{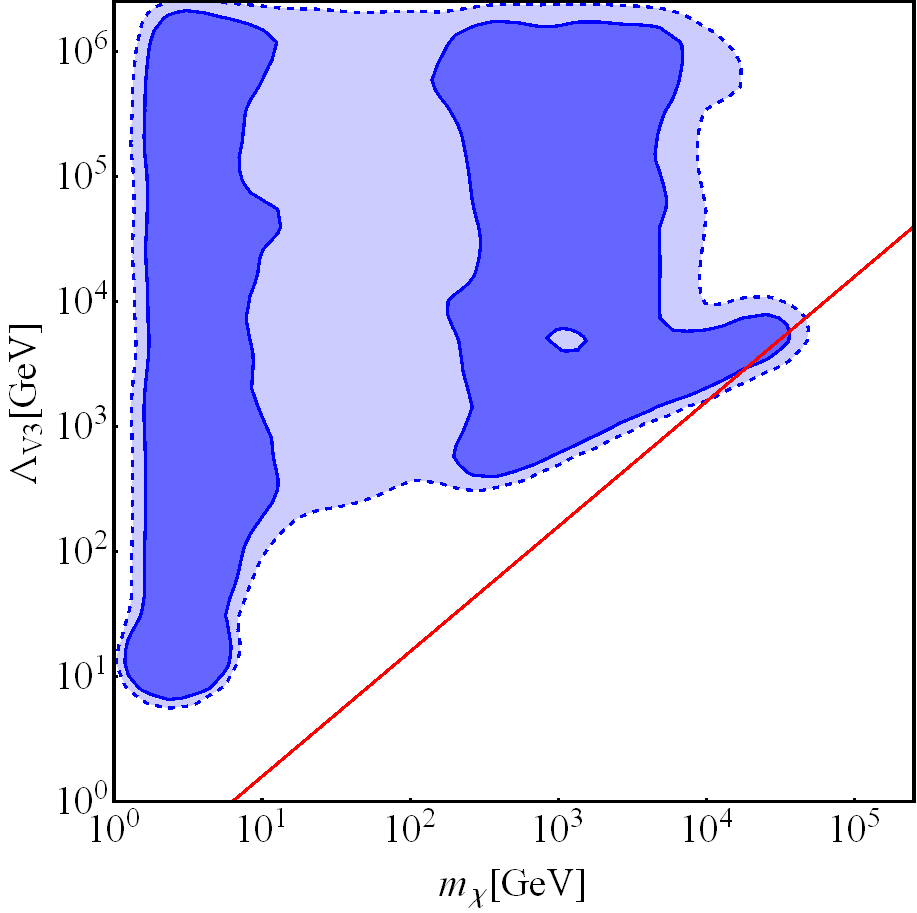}
\includegraphics[width=5.4cm]{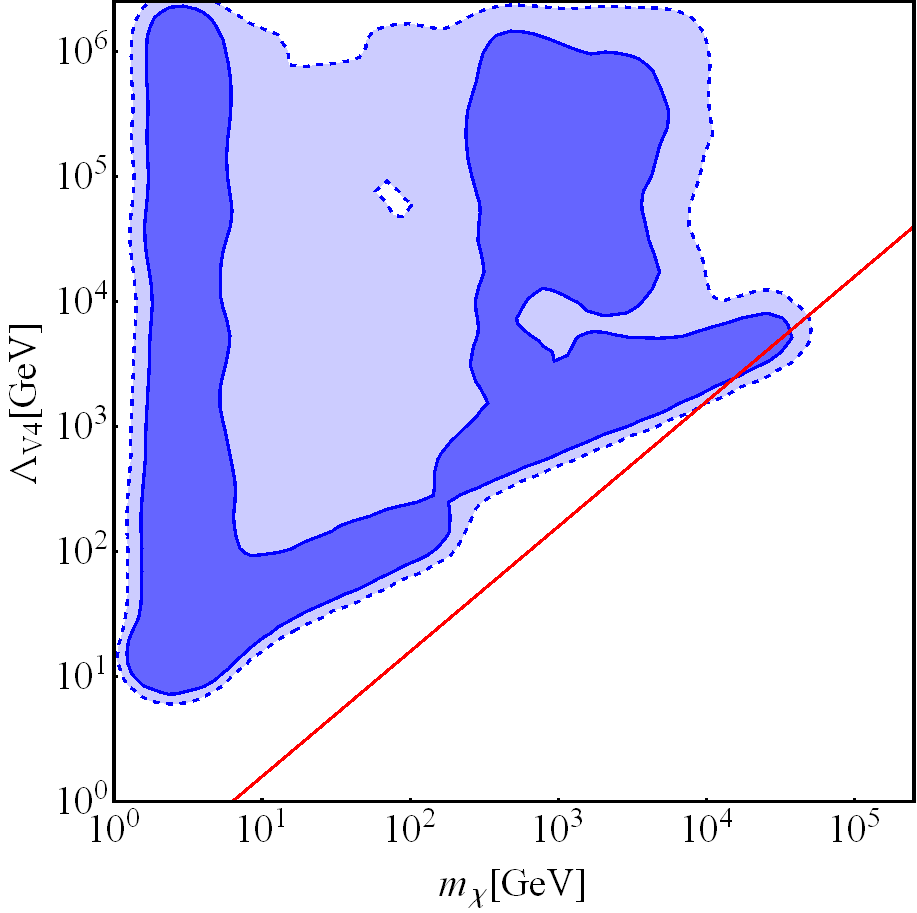}
\caption{Posterior probability distribution marginalized to the $\Lambda_{Vi}$ scale ($i=1-4$) and the mass of the vector boson DM particle.}
\label{figLambdaVBall}
\end{figure}


\clearpage
\bibliography{EFTbibtex}

\end{document}